# A statistical-mechanical framework for mechanically adaptive cytoskeletal organization

**Yuika Ueda[1] and Shinji Deguchi[1,2,3,†]**

[1] Division of Bioengineering, Graduate School of Engineering Science, University of Osaka

[2] Global Center for Medical Engineering and Informatics, University of Osaka

[3] R$^3$ Institute for Newly-Emerging Science Design, University of Osaka

† Corresponding author

Address: 1-3 Machikane-yama, Toyonaka, Osaka 560-8531, Japan

E-mail: deguchi.shinji.es@osaka-u.ac.jp

Phone: +81 6 6850 6215

ORCID: 0000-0002-0556-4599

## Abstract

Living cells continuously remodel their cytoskeleton in response to mechanical cues. Although these responses have been extensively documented, it remains unclear why continuous changes in the mechanical environment give rise to distinct intracellular architectures rather than gradual structural variation. Here, we introduce a statistical-mechanical framework in which alternative cytoskeletal organizations are represented as ensembles of microscopic configurations, allowing configurational entropy to compete with mechanically dependent interaction energies. Rather than reproducing the full molecular complexity of the cytoskeleton, the model asks which features of mechanically adaptive organization emerge from this minimal physical description. The framework predicts three successive structural transitions corresponding to stress fiber formation, alignment, and lateral aggregation. When these transitions are placed on a common cellular-tension axis that increases with substrate stiffness, the predicted sequence is consistent with our measurements of correlation length and anisotropy in senescent fibroblasts. The preservation of this stiffness-dependent sequence despite altered cellular physiology suggests that the observed ordering reflects a robust physical principle rather than a cell-state-specific phenomenon. Together, these results establish a statistical-mechanical framework for understanding how continuous mechanical cues bias the statistical selection of distinct cytoskeletal architectures.

Keywords: statistical mechanics, configurational entropy, mechanobiology, cytoskeletal architecture, mechanoadaptation

## Introduction

Living cells continuously remodel their intracellular architecture in response to changes in their mechanical environment. A characteristic example is the actin cytoskeleton, whose organization is strongly regulated by extracellular matrix stiffness (1). As substrate stiffness increases, stress fibers (SFs) emerge, subsequently align in a common direction, and eventually assemble into thick lateral bundles (2–5). Similar sequential reorganizations are also observed during cell spreading on sufficiently stiff substrates (6–8). These observations indicate that continuous mechanical cues can give rise to qualitatively distinct cytoskeletal architectures rather than gradual changes in structural organization.

Although considerable progress has been made in identifying the molecular pathways that mediate mechanotransduction, including mechanisms underlying the stepwise progression of cytoskeletal reorganization (3), how mechanical cues bias the selection of distinct cytoskeletal organizations remains incompletely understood. Elucidating this state-selection process is particularly important because substrate stiffness regulates fundamental cellular functions, including proliferation, differentiation, and apoptosis (9,10).

Statistical mechanics provides a complementary framework for addressing this problem. Rather than describing every molecular interaction individually, it explains macroscopic order by considering the statistical properties of microscopic configurations. Statistical-mechanical concepts have long been central to understanding phase transitions in condensed matter physics (11) and, more recently, have been successfully extended to biological phenomena such as biomolecular phase separation (12). Here we apply this perspective to intracellular cytoskeletal organization by explicitly counting microscopic configurations underlying distinct states of SF organization. The corresponding configurational entropy is balanced against mechanically dependent interaction energies to determine which structural state becomes statistically favored. We ask how far mechanically adaptive cytoskeletal organization can be understood within this minimal statistical-mechanical framework.

To motivate such a physical treatment, we first examined whether the stiffness-dependent sequence of SF reorganization is preserved in senescent fibroblasts. Despite marked reductions in migration and metabolic activity, senescent cells exhibited a sequence of structural changes consistent with that previously reported in proliferating cells and cell lines. Although this observation does not isolate a purely physical mechanism, because mechanotransduction remains active in senescent cells, it suggests that the ordering sequence is preserved across substantially different cellular states and therefore motivates a coarse-grained physical description.

Based on this perspective, we develop statistical-mechanical models describing successive transitions in SF formation, alignment, and lateral aggregation. The three transitions are represented on a common cellular-tension axis that increases with substrate stiffness, allowing direct comparison with experimental observations. Molecular regulators are incorporated implicitly through effective physical parameters rather than modeled individually. We then compare the theoretical predictions with stiffness-dependent cytoskeletal reorganization in senescent fibroblasts and show that the predicted ordering is preserved despite reduced migratory and metabolic activity. Together, these results provide a statistical-mechanical framework for understanding how continuous mechanical cues bias the selection of distinct cytoskeletal architectures. This manuscript is based in part on the doctoral dissertation of Y.U. (13).

## Experimental methods

Primary human foreskin fibroblasts (HFF-1, ATCC) were cultured in high-glucose Dulbecco's modified Eagle's medium (Wako) supplemented with 15% fetal bovine serum (Sigma-Aldrich) and 1% penicillin–streptomycin (Wako). Cells were maintained at 37 °C with 5% $CO_2$ and passaged at a 1:3 ratio upon reaching 70–80% confluence. Cells at passages 20–30, which exhibit advanced senescent characteristics accompanied by limited metabolic and migratory activity and pronounced SF development (14), were used for experiments. Polyacrylamide (PA) hydrogels of defined stiffness were prepared as previously described (15). Briefly, acrylamide/bis-acrylamide solutions (5–8% acrylamide, 0.1–0.48% bis) were polymerized on glass coverslips, functionalized using a Sulfo-SANPAH cross-linker (Thermo Fisher Scientific) under UV activation, and coated with fibronectin (10 $\mu$g/ml, Sigma-Aldrich). Gel stiffness was quantified by atomic force microscopy (NanoWizard II; qp-BioAC-Cl-10 cantilevers, spring constant 0.06–0.18 N/m, NANOSENSORS), in which force–indentation curves were collected to determine the elastic modulus (7 kPa, 20 kPa, and 41 kPa) using the Hertz model from at least three independent gel preparations. For staining, cells grown on glass coverslips or PA hydrogels were allowed to spread for a fixed duration and were fixed with 4% paraformaldehyde, permeabilized with 0.1% (v/v) Triton X-100, and blocked with 10% (w/v) normal goat serum in PBS. F-actin was visualized using Alexa Fluor 488–conjugated phalloidin (Thermo Fisher Scientific). Images were acquired using a confocal microscope (FV3000, Olympus; UPlanApo 60× oil-immersion objective, NA 1.42). For each substrate stiffness, 10–15 cells were analyzed.

Actin organization was quantified by combining structure tensor-based local orientation mapping with spatial autocorrelation analysis to assess filament formation. Individual cells were manually segmented, and two-dimensional fast Fourier transform-based autocorrelation maps were computed in ImageJ/Fiji. Radial averaging yielded one-dimensional autocorrelation functions $g(r)$, where $r$ is the radial distance, which were normalized by their value at $r = 0$. The decay of $g(r)$ was fitted with an exponential function, $\exp(-r/\xi)$, from which correlation length $\xi$ was obtained as a measure of actin structural correlation. Directional ordering was quantified using anisotropy analysis with the FibrilTool plugin, which computes values between 0 (random) and 1 (perfect alignment) based on local intensity gradients. Mean anisotropy values were calculated for each cell to enable comparison across substrate stiffness conditions. Statistical significance was evaluated using the Mann-Whitney *U*-test.

## Model overview

Cells exhibit a sequence of structural transitions in actin organization as substrate stiffness increases (Fig. 1) (3). In response to increasing stiffness, cells progress through four qualitatively distinct states: i) a disordered phase in which actin particles are randomly distributed; ii) a bundled phase in which these particles aggregate into SFs with random orientations; iii) an aligned phase in which SFs acquire a common direction; and iv) an aggregated phase in which aligned SFs coalesce laterally. This sequence of state changes motivates a theoretical description in terms of a hierarchy of physical phase transitions. To explore the fundamental principles governing these transitions, we construct a theoretical framework based on statistical mechanics. Unlike conventional phase transition theories based on a single order parameter, the present framework describes a hierarchy of distinct orders, each emerging at a different stiffness threshold and inherited by subsequent phases. To analyze this hierarchy, we develop distinct theoretical models for each of the three phase transitions, SF formation (Phase transition I, PT1), directional alignment (Phase transition II, PT2), and fiber aggregation (Phase transition III, PT3), each capturing a different level of structural organization. The models share a common physical basis but differ in formulation according to the dominant order in each state. This framework is intended as a minimal, conceptual model that isolates the essential energetic and entropic relationships, rather than as a complete thermodynamic derivation of cytoskeletal organization. Because actin architecture varies across a cell, different regions can occupy different states at the same time. We therefore partition the cell into local regions and treat each region as an isolated lattice, that is, as a single coarse-grained sub-system with a side length of a rather than the entire cell (Fig. 1b). Each transition is then analyzed within one such sub-system, and the coexistence of different states in different regions is consistent with each region undergoing its own local transition. This viewpoint keeps the microstate counting tractable while allowing the cell as a whole to display a mixture of states.

For each transition we define an order parameter, construct the Helmholtz free energy from the energy and the configurational entropy, and locate the state that minimizes it. From these free energies we also determine the nature of each transition, that is, whether the order parameter changes continuously or discontinuously as its control parameter crosses the relevant threshold. In the three sections that follow, each control parameter is treated separately: the sustained tension for PT1 and PT2, and a binding-related parameter for PT3. After the three transitions are established individually, we express all three control parameters as increasing functions of a single cellular tension that rises with stiffness, which fixes the order in which the transitions occur. The predicted order is then compared with the measured stiffness dependence of actin organization in the comparison section, and the free-energy landscapes are brought together in the Discussion (Fig. 6). Numerous studies have shown that increasing substrate stiffness monotonically increases actomyosin-generated tension through focal adhesions. We therefore treat substrate stiffness as an experimentally controllable proxy for sustained intracellular tension, which serves as the control parameter in our theoretical framework.

## Phase transition 1 – Formation of stress fibers as a physical phase transition

We begin by considering the phase transition process, in which randomly distributed actin particles form SFs, characteristic of PT1. Here, actin particles denote coarse-grained actin-based structural units that represent material progressing toward the next structural state, ranging from monomeric actin that polymerizes into filaments to filament segments that assemble into SFs. The cell is represented as a two-dimensional square with a side length of $a$, on which actin particles are distributed (Fig. 2a), where $a$ denotes the side length measured in lattice units, with the lattice spacing set to unity; numerically, $a$ also corresponds to the number of lattice sites per side. We focus on spontaneous phenomena that emerge independently of macroscopic changes in cell shape or volume, which are not included in the present model. We also assume that the temperature of the cell remains constant. The Helmholtz free energy of the cell is defined as follows:

$$A = E - TS \qquad (1-1)$$

where $E$, $T$, and $S$ represent energy, temperature, and entropy, respectively. We derive the specific form of Eq. (1-1) by considering the distribution of actin particles. The square cell is subdivided into smaller units referred to as a lattice. The total number of lattice sites is equal to the area of the cell $V$:

$$V = a^2. \qquad (1-2)$$

Let $N_a$ represent the total number of actin particles within the cell, which remains constant as actin works as a "housekeeping" protein. The cell is sufficiently large relative to the area occupied by all actin particles to ensure $N_a < V$, and each lattice site can be occupied by at most one actin particle.

We define SFs as the arrangement of actin particles that align in a straight line from one edge of the cell to the other (Fig. 2a, red). Any other configurations are considered to be randomly distributed actin particles (Fig. 2a, green). Although SFs in actual cells may not always form perfect straight lines, we define SFs in this way to enable a strict count of the possible microscopic states, or microstates, of this macroscopic cell system. Here note that, in statistical mechanics systems such as those applied to biological contexts (16–20), the distinction between macroscopic and microscopic variables is not simply determined by physical size. We take the length $a$ as the minimum length required to count as an SF; cases with SFs of varying length or shape can be treated by interpreting the present system as a sub-system of a larger cellular domain, which is beyond the present scope.

Let $L_s$ represent the number of SFs formed by actin particles. To characterize the phase state, we define an order parameter, which is a macroscopic variable that determines the degree of order within the cell, based on the extent to which SFs are formed. Specifically, the order parameter $x_1$ is defined as the ratio of actin particles involved in SF formation to the total number:

$$x_1 = \frac{L_s a}{N_a} \tag{1-3}$$

When $x_1$ = 0, no SFs are formed, indicating a completely disordered state. Meanwhile, when $x_1$ = 1, all actin particles form SFs, representing a fully ordered state within the cell. We determine the specific form of the entropy by counting the possible microstates. Actin particles can exist in two distinct states: as part of non-SFs or as part of SFs. First, the number of possible microstates of actin particles or non-SFs (Fig. 2a, green) is given by:

$$W_p = \frac{(V - L_s a)!}{(N_a - L_s a)!\,(V - N_a)!} \tag{1-4}$$

where the available lattice sites are those not involved in SF formation (i.e., $V - L_s a$), and the number of available actin particles and empty lattice sites are $N_a - L_s a$ and $V - N_a$, respectively. Similarly, the number of possible microstates of SFs (Fig. 2a, red) is given by

$$W_s = \frac{2a!}{L_s!\,(2a - L_s)!}, \qquad (1-5)$$

in which the number of possible configurations is calculated for arranging a total of $L_s$ SFs. We do not distinguish between SF configurations that are symmetrical with respect to the starting position, either vertically or horizontally. The number of possible microstates is thus calculated within the range of $a + a = 2a$. Taken together, the system entropy is described, with the Stirling approximation, by

$$S_1 = k_b \ln W_p W_s$$
$$= k_b \begin{Bmatrix} (V - N_a x_1)\ln(V - N_a x_1) - N_a(1 - x_1)\ln N_a(1 - x_1) - (V - N_a)\ln(V - N_a) \\ +4a\ln a - 2a\ln 2 - \frac{N_a}{a} x_1 \ln N_a x_1 - \left(2a - \frac{N_a}{a} x_1\right)\ln(2V - N_a x_1) \end{Bmatrix} \qquad (1-6)$$

where $k_b$ is the Boltzmann constant (Fig. S1). Equation (1-6) follows directly from the microstate counting in Eqs. (1-4) and (1-5); its derivative reproduces the transition condition given below, confirming the consistency between the free-energy landscape and the transition curve.

Next, to describe the energy $E_1$, we define the single-particle energy of an actin particle as $\mu_1^0$ and the interaction energy due to binding as $\mu_1^s$, respectively; $\mu_1^0$ is an intrinsic energy that reflects the inherent properties of the actin particle and is considered a constant, while $\mu_1^s$ represents the energy associated with the stability of the binding and is generally a function that can vary depending on chemical and physical influences from the surrounding environment. A stiffer substrate allows cells to generate larger traction forces due to the law of action and reaction (10), and given that actin–myosin complexes including SFs become more stable under such increased tension, a property known as a catch bond (21–25), the interaction energy is expressed as a function of tension $F_1$:

$$\mu_1^s(F_1) = B\exp(-CF_1) \qquad (1-7)$$

so that larger tension lowers the interaction energy and thus stabilizes the bound (SF) state. Here, $B$ is the interaction energy due to binding in the absence of tension, and $C$ determines the effect of $F_1$. Because Eq. (1-7) makes the bound state more favorable as $F_1$ grows, it captures catch-bond-like stabilization within the present energetic description. The total energy of the cell is expressed using the number of non-SF actin particles, $N_a - L_s a$, and the number of particles forming SFs, $L_s$ (Fig. S2):

$$E_1 = \mu_1^0 (N_a - L_s a) + \mu_1^s L_s a \tag{1-8}$$

From Eqs. (1-6) and (1-8), the free energy of Eq. (1-1) is expressed as

$$\begin{gathered} A_1 = \mu_1^0 (N_a - x_1 N_a) + B N_a x_1 \exp(-C F_1) \\ -T k_b \begin{Bmatrix} (V - N_a x_1) \ln(V - N_a x_1) - N_a (1 - x_1) \ln N_a (1 - x_1) - (V - N_a) \ln(V - N_a) \\ +4 a \ln a - 2 a \ln 2 - \frac{N_a}{a} x_1 \ln N_a x_1 - \left(2a - \frac{N_a}{a} x_1\right) \ln(2V - N_a x_1) \end{Bmatrix}. \end{gathered} \tag{1-9}$$

To derive the conditions for PT1, we take the partial derivative of Eq. (1-9) with respect to $x_1$:

$$\frac{\partial A_1}{\partial x_1} = N_a \left[ (\mu_1^s - \mu_1^0) - k_b T \left\{ - \ln(V - N_a x_1) + \ln N_a (1 - x_1) - \frac{1}{a} \ln N_a x_1 + \frac{1}{a} \ln(2V - N a x_1) \right\} \right] \tag{1-10}$$

The order parameter $x_1^*$ that satisfies $\partial A_1 / \partial x_1 = 0$ is

$$x_1^* \approx 1 - \frac{V}{N_a} \exp \frac{\mu_1^s - \mu_1^0}{k_b T}, \tag{1-11}$$

in which the approximation $V \gg N_a$ is used. The free energy has an extremum in $0 < x_1^* < 1$, indicating that a symmetry-breaking phase transition is induced at $x_1^* \neq 0$. Thus, the criterion for PT1 is given by

$$\mu_1^s - \mu_1^0 < k_b T \ln \frac{N_a}{V}, \tag{1-12}$$

suggesting that if the interaction energy $\mu_1^s$ is low, the cell system becomes stable by forming SFs and establishing order. In other words, the system spontaneously moves towards an ordered state characterized by SF formation when substrate stiffness is high. Conversely, if $\mu_1^s$ is high, the cell is more stable when actin particles remain randomly distributed without forming SFs. While this model assumes constant cell volume and actin particle number, the result suggests that a higher effective concentration of actin particles, $N_a / V$, facilitates SF formation. Interestingly, even though this model does not explicitly consider the effect of concentration, its impact on the cell stability is revealed within the framework of statistical mechanics.

Using the more rigorous Eq. (1-10), we numerically show the relationship between the order parameter $x_1$ and the free energy $A_1$ (Figs. 2b-2g). Hereafter, unless otherwise stated, these parameters are used: $N_a$ = 200 (corresponding to 20 actin particles per filament in the calculation), $T$ = 309.5 K, $k_b$ = 1.38 × 10^-23 J/K, $F_1$ = 1 × 10^-12, $a$ = 20, $B$ = 1.5 × 10^-20, and $C$ = 2.5 × 10^12. $B$ and $C$, which affect the magnitude of the energy, are determined given that the binding free energy of proteins generally falls within the range of 1–10 kcal/mol (26). These are biologically plausible orders of magnitude rather than fitted values; because the model is conceptual and not intended to reproduce absolute measurements, this choice places the transitions in realistic cellular regimes rather than at arbitrary points. A table of all parameters, with units and physical meaning, is given in the Supplementary Information (Table S1). Under conditions where the tension $F_1$ is small, the free energy $A_1$ reaches its minimum at $x_1$ = 0 (Fig. 2b, blue and orange), indicating that the cell system is stable in a disordered state where actin particles are randomly distributed without forming SFs. Once $F_1$ exceeds a certain threshold described by Eqs. (1-7) and (1-11), $A_1$ shifts to its minimum at higher values of $x_1$ (Fig. 2b, green and red), suggesting that in environments with larger forces, such as those with greater substrate stiffness, the cell system is more stable as it adopts an ordered phase characterized by the presence of one or more SFs. In environments where the total number of effective actin particles $N_a$ is small, the free energy $A_1$ reaches its minimum at $x_1$ = 0 (Fig. 2e, blue), suggesting that the cell system is stable in a state where actin particles are randomly distributed. As $N_a$ increases, in contrast, the free energy minimum shifts to a larger value of $x_1$ (Fig. 2ee, green and red), indicating that cells with higher actin concentrations are more stable in an ordered phase where SFs are formed (Figs. 2f and 2g). The analytical results for the order parameter $x_1^*$, at which the free energy is minimized, also demonstrate a similar phase transition as in the numerical analysis (data not shown).

To determine the nature of PT1, we examine how $x_1^*$ behaves as $F_1$ crosses the threshold and how the free-energy landscape changes there. Across the threshold, $A_1(x_1)$ retains a single minimum whose location moves continuously from $x_1 = 0$ to positive values (Fig. S3), and $x_1^*$ rises continuously from zero rather than jumping to a finite value. There is no second competing minimum and no hysteresis. PT1 is therefore a continuous transition. Consistent with this, expanding $A_1$ near $x_1 = 0$ gives a leading quadratic term whose coefficient changes sign at the threshold, so the onset is second-order-like; the expansion is given in the Supplementary Information. We further checked that the transition persists over a broad range of the parameters $B$, $C$, $\mu_1^0$, and $N_a$, with the threshold shifting smoothly as each is varied (Fig. S4), which shows that the result does not depend on a particular choice of values.

## Phase transition II – Directional alignment of stress fibers as a physical phase transition

We explore the mechanism of the phase transition from a cell structure with SFs formed in random directions to one where SFs acquire a specific orientation, characteristic of Phase transition II (PT2). In the previous phase, the presence of SFs alone was regarded as an ordered state. In contrast, PT2 is interpreted as the emergence of a new order, where SFs align in a particular direction. In this new phase, SFs align either vertically or horizontally (Fig. 3a). Although in actual cells not all actin particles form SFs, we consider that the cell is entirely in an SF-forming phase to focus on the essential aspects of this specific phase transition. In addition, while actual SFs can align in various directions, we count only those aligning vertically or horizontally. While our current model focuses on the actin cytoskeleton within a simplified two-dimensional lattice, it can be extended to include additional cellular components, higher-dimensional lattices, or curved SF arrangements. These extensions would increase the complexity of energetic and entropic terms while retaining the core theoretical structure. In particular, cases involving SFs of varying lengths, shapes, or orientations can be addressed by interpreting the current model as a sub-system within a larger cellular domain.

Within the square lattice framework as used for PT1, let us consider the case where the total number of SFs is $n_2$. We denote the number of SFs aligning along the $x$-axis as $n_2^x$ (Fig. 3a, green) and those aligning along the $y$-axis as $n_2^y$ (red), and thus

$$n_2 = n_2^x + n_2^y \qquad (2-1)$$

The new order parameter that determines the phase state here is defined as follows to evaluate the extent to which SFs are aligned in the same specific directions.

$$x_2 = \frac{n_2^x - n_2^y}{n_2} \qquad (2-2)$$

When $x_2$ = 0, SFs are evenly distributed between the two axes, indicating a disordered state; meanwhile, when $x_2$ = 1 or $x_2$ = -1, SFs are aligned only along either the $x$-axis or $y$-axis, reflecting a state with directional order. The two signs of $x_2$ correspond to the two equivalent alignment directions, which will be important for the nature of the transition. The number of possible microstates determined from the distribution of SFs is described by

$$W = \frac{n_2!}{n_2^x!\, n_2^y!}, \tag{2-3}$$

and then the entropy is described, with the Stirling approximation, by

$$\begin{aligned} S_2 &= k_b \ln W \\ &= k_b \left( n_2 \ln n_2 - \frac{1+x_2}{2} \ln\left(\frac{1+x_2}{2} n_2\right) - \frac{1-x_2}{2} \ln\left(\frac{1-x_2}{2} n_2\right) \right) \end{aligned} \tag{2-4}$$

(Fig. S5). We define the energy of a single SF bearing a tension of $F_2$ based on the catch-bond properties:

$$\mu_2(F_2) = \mu_2^0 \exp\left(-\frac{F_2}{F_2^0}\right), \tag{2-5}$$

in which $F_2$ (= $f_2 a$) is the maximum tension that fully formed single SFs can sustain; $f_2$ is the unit tension, $F_2^0$ is a normalizing factor, and $a$ is the cell edge length as in PT1. We express the per-fiber tension as a function of the unit tension $f_2$ and the number of interfering crossings in a dimensionally consistent form. Tension in proliferative cells is known to be produced by the interaction between ubiquitous nonmuscle actin and myosin II filaments, which occurs when these proteins align in a straight configuration (27,28). Based on this, we assume that when one SF crosses another perpendicularly, the intersection point lacks the linear structure effective for tension generation and maintenance. If an SF does not experience such interference from others, the tension it can generate reaches its maximum value of $F_2$. The tensions in the $x$-axis and $y$-axis directions are then given as follows:

$$\begin{aligned} F_2^y &= f_2(a - n_2^x) \\ F_2^x &= f_2(a - n_2^y) \end{aligned} \tag{2-6}$$

From Eqs. (2-5) and (2-6), the energy $E_2$ and free energy $A_2$ are given by:

$$E_2 = n_2^y \mu_2(F_2^y) + n_2^x \mu_2(F_2^x) \tag{2-7}$$

(Fig. S6) and

$$A_2 = \mu_2^0 \left\{ \frac{1-x_2}{2} n_2 \mu_2^0 \exp\left( -\frac{f_2\left(a - \frac{(1+x_2)n_2}{2}\right)}{F_2^0} \right) + \frac{1+x_2}{2} n_2 \mu_2^0 \exp\left( -\frac{f_2\left(a - \frac{(1-x_2)n_2}{2}\right)}{F_2^0} \right) \right\}$$

$$-Tk_b \left( n_2 \ln n_2 - \frac{1+x_2}{2} \ln\left(\frac{1+x_2}{2} n_2\right) - \frac{1-x_2}{2} \ln\left(\frac{1-x_2}{2} n_2\right) \right), \qquad (2-8)$$

respectively. To analyze the critical condition, taking the second derivative of Eq. (2-8) with respect to $x_2$ and substituting $x_2 = 0$ yield

$$\frac{\partial^2 A_2(0)}{\partial x_2^2} = \frac{n_2^2 f_2 \mu_2^0}{4F_2^0} \left( -4 + \frac{f_2 n_2}{F_2^0} \right) \exp\left( -\frac{f_2\left(a - \frac{n_2}{2}\right)}{F_2^0} \right) + n_2 T k_b. \qquad (2-9)$$

With the approximation $\exp\left(-f_2\left(a - \frac{n_2}{2}\right)/F_2^0\right) \approx 1$ and solving $\partial^2 A_2(0)/\partial x_2^2 = 0$,

$$f_2^* = \frac{2\left( F_2^0 \mu_2^0 n_2 - F_2^0 \sqrt{{\mu_2^0}^2 n_2^2 - k_b \mu_2^0 n_2^2 T} \right)}{\mu_0 n^2}, \qquad (2-10)$$

and thus the cell system stabilizes at $x_2 = 0$ when $f_2 < f_2^*$, and at $x_2 \neq 0$ when $f_2 > f_2^*$.

Using the rigorous Eq. (2-8) with $\mu_2^0 = 10^{-19}$, $F_2^0 = 10^{-9}$, and $n_2 = 10$ (i.e., 10 fiber units), we visualized the relationship between the free energy $A_2$ and the order parameter $x_2$ (Fig. 3b, bule), showing that $A_2$ reaches its minimum at $x_2 = 0$ when the force $f_2$ is small. Meanwhile, the free energy has a convex shape and reaches its minimum at $x_2 \neq 0$ when the force is large (Fig. 3b, green). Thus, SFs sustaining low tension remain stable with random orientation, but once the tension exceeds a certain threshold the influence of energy becomes more dominant than entropy, leading to a phase where SFs align in a consistent direction (Fig. 3c, d). When the total number of distinct SFs, $n_2$, is small, the order parameter corresponding to the minimum free energy begins to change at higher force levels (Fig. 3e). In contrast, when $n_2$ is large, the change in the order parameter starts at lower force levels, suggesting that PT2 occurs more easily with more SFs under the same force conditions than with fewer SFs.

PT2 is a continuous, symmetry-breaking transition of pitchfork type. The free energy is symmetric in $x_2$, so at the threshold the single minimum at $x_2 = 0$ does not jump; instead it becomes unstable and is replaced by two equivalent minima at $x_2 = +/-x_2^*$ that grow continuously from zero (Fig. S7). The magnitude $|x_2^*|$ rises from zero above $f_2^*$ without a discontinuity or hysteresis, and the two branches reflect the equivalence of the two alignment directions. This behavior, a sign change of the curvature at $x_2 = 0$ followed by symmetric minima, is characteristic of a second-order (pitchfork) transition, in contrast to a first-order jump. We confirmed that the transition persists as $\mu_2^0$, $F_2^0$, a, and $n_2$ are varied over broad ranges, with the threshold shifting smoothly (Fig. S8). The full expressions and the sign analysis are given in the Supplementary Information.

## Phase transition III – Lateral aggregation of aligned stress fibers as a physical phase transition

Next, we focus on the mechanism of Phase transition III (PT3) in which SFs, after acquiring directional alignment in PT2, undergo further transformation to undergo interfiber coupling. As in the previous sections, we consider a simple model where directionally aligned SFs are distributed within a square cell lattice of side length $a$ with constant volume and temperature (Fig. 4a). In actual cells, not all SFs align in a single direction, but we focus on the key aspects of how directionally aligned SFs cluster after PT2. Let the total number of SFs be $n_3$ , and let $n_3^{SF}$ and $n_3^{ass}$ represent the number of independent and assembled ones, respectively:

$$n_3 = n_3^{SF} + n_3^{ass} \tag{3 − 1}$$

We consider the case where a single aggregate is present. Although multiple aggregates form in actual cells, this complexity can be addressed by adjusting entropy calculations. Here, however, the focus is on establishing a framework for analyzing the fundamental principles and significance of PT3. The order parameter is defined as follows to evaluate the extent to which SFs form aggregates:

$$x_3 = \frac{n_3^{ass}}{n_3} \tag{3 − 2}$$

SFs are independently distributed at $x_3$ = 0, reflecting a disordered state, while they undergo interfiber coalescence at $x_3$ = 1. The number of microstates is expressed as follows, focusing on the distribution patterns:

$$W = \frac{(a + 1 - n_3^{ass})!}{(n_3 - n_3^{ass})!\,(a - n_3^{ass})!} \tag{3 − 3}$$

The entropy is described, with the Stirling approximation, by

$$S_3 = k_b \ln W$$
$$= k_b\{(a+1-n_3x_3)\ln(a+1-n_3x_3) - n_3(1-x_3)\ln n_3(1-x_3) - (a-n_3)\ln(a-n_3)\} \quad (3-4)$$

(Fig. S9). The energy of individual SFs is given by Eq. (2-5) as used in PT2,

$$\mu_3 = \mu_3^0 \exp\left(-\frac{cf_3}{F_3^0}\right), \quad (3-5)$$

but the interpretation here is that $c$ is a concentration-dependent parameter of binding proteins such as α-actinin that stabilize the connections between SFs. The energy and free energy are then be expressed by

$$E_3 = n_3\left\{(1-x_3)\mu_3^0 \exp\left(-\frac{c_{SF}f_3}{F_3^0}\right) + x_3\mu_3^0 \exp\left(-\frac{c_{ass}f_3}{F_3^0}\right)\right\} \quad (3-6)$$

(Fig. S10), and

$$A_3 = n_3\left\{(1-x_3)\mu_3^0 \exp\left(-\frac{c_{SF}f_3}{F_3^0}\right) + x_3\mu_3^0 \exp\left(-\frac{c_{ass}f_3}{F_3^0}\right)\right\}$$
$$-Tk_b\{(a+1-n_3x_3)\ln(a+1-n_3x_3) - n_3(1-x_3)\ln n_3(1-x_3) - (a-n_3)\ln(a-n_3)\}, \quad (3-7)$$

respectively, where $c_{SF}$ and $c_{ass}$ are $c$ for independently distributed SFs and assembled SFs, respectively. Differentiating Eq. (3-7) with respect to $x_3$ is

$$\frac{\partial A_3}{\partial x_3} = n_3\left(-\mu_3^0 \exp\left(-\frac{c_{ass}f_3}{F_3^0}\right) + \mu_3^0 \exp\left(-\frac{c_{SF}f_3}{F_3^0}\right)\right) - Tk_b(\ln(a+1-n_3x_3) + \ln(n_3-n_3x_3)), \quad (3-8)$$

and the order parameter at $\partial A_3/\partial x_3 = 0$ is expressed as

$$x_3^* = \frac{(a+1)\exp\left(\dfrac{-\mu_3^0 \exp\left(-\frac{c_{ass}f_3}{F_3^0}\right) + \mu_3^0 \exp\left(-\frac{c_{SF}f_3}{F_3^0}\right)}{Tk_b}\right) - n_3}{n_3\left(\exp\left(\dfrac{-\mu_3^0 \exp\left(-\frac{c_{ass}f_3}{F_3^0}\right) + \mu_3^0 \exp\left(-\frac{c_{SF}f_3}{F_3^0}\right)}{Tk_b}\right) - 1\right)}. \quad (3-9)$$

The free energy reaches an extremum and stabilizes at this specific value $x_3^*$ in the range $0 < x_3^* < 1$, as long as $x_3^*$ is not equal to 0. The condition for PT3 is now described by

$$-\mu_3^0 \exp\left(-\frac{c_{ass} f_3}{F_3^0}\right) + \mu_3^0 \exp\left(-\frac{c_{SF} f_3}{F_3^0}\right) < Tk_b \ln\frac{n_3}{a+1}, \quad (3-10)$$

indicating that even when the energy difference on the left side remains unchanged, PT3 can still occur if the total number of SFs is large.

Visualization of PT3 with $n_3$ = 10 (i.e., 10 fiber units) and $F_3^0 = 1 \times 10^{-9}$ shows that entropy dominates when the tension is weak with a low binding protein concentration, leading to a state where the free energy is minimized at $x_3$ = 0 (Fig. 4b, blue and orange). On the other hand, a high binding protein concentration leads to stabilization of the cell system with SF aggregates (Fig. 4b, green). One straightforward interpretation of the parameter $c_{ass}$ is the concentration of actin binding proteins that allow SFs to engage in interfiber coupling, but another is that $c_{ass}$ also represents the internal and external factors that drive changes in the size of SFs through the enlargement of focal adhesions and consequently increase the sustained tension, which is also enhanced by substrate stiffness (Fig. 4c, d) (29–33). The threshold concentration decreases as the total number of SFs increases (Fig. 4e). This observation is consistent with the theoretically derived phase transition conditions (Eqs. (3-9) and (3-10)). Free-energy curves and maps for the second control variable $n_3$ are shown in Fig. 4f-h.

PT3 is a continuous transition. Across the threshold set by Eq. (3-10), $A_3(x_3)$ keeps a single minimum whose location moves continuously from $x_3 = 0$ to positive values, and $x_3^*$ rises from zero without a discontinuity, a second minimum, or hysteresis (Fig. S11). The onset is therefore second-order-like rather than a first-order jump. We confirmed that the transition persists as $\mu_3^0$, $F_3^0$, $f_3$, and $n_3$ are varied over broad ranges, with the threshold shifting smoothly (Fig. S12). Details are given in the Supplementary Information.

## Stiffness-dependent reorganization compared with the model

Having established the three transitions and their order in the model, we test the predictions against measurements of actin organization across substrate stiffness. HFF-1 cells in an advanced senescent state were cultured on substrates of increasing rigidity (7, 20, and 41 kPa) and on glass. F-actin staining revealed a sequential change in intracellular organization with increasing stiffness (Fig. 5a). Spatial autocorrelation analysis showed a significant increase in correlation length $x_i$ from 7 to 20 kPa, indicating the emergence of spatially extended filament continuity (Fig. 5c), whereas anisotropy increased significantly on stiffer substrates, from 20 to 41 kPa, reflecting the onset of directional alignment (Fig. 5d). On glass, both quantities increased further (Fig. 5c, 5d), consistent with orientation maps showing progressively ordered actin (Fig. 5a). Adjacent-condition significance was assessed with the Mann-Whitney test. These measurements resolve three structurally distinct states that emerge in a fixed, threshold-like sequence as stiffness increases, in line with the stepwise ordering the model predicts. Establishing the finer signatures of a transition, such as criticality or hysteresis, would require a denser range of stiffnesses; the present data are well suited to testing the predicted sequence, which is the comparison we pursue here.

The two measured quantities serve as experimental readouts of the modeled transitions. SF formation (PT1) converts randomly distributed actin into extended, continuous filaments, raising xi; directional alignment (PT2) reorients fibers toward a common axis, raising anisotropy; lateral aggregation (PT3) consolidates aligned fibers into thicker bundles, raising $x_i$ further at the highest stiffness. Reading $x_i$ as the observable for PT1 and PT3 and anisotropy as that for PT2, and normalizing each observable to its own range, the two quantities rise at different stiffness ranges rather than together: $x_i$ first (7 to 20 kPa), anisotropy later (20 to 41 kPa), and both on glass (Fig. 5e). In the model, expressing the three control parameters as increasing functions of a single cellular tension that rises with stiffness places the transitions in a fixed sequence from formation to alignment to aggregation (Fig. 5e): formation precedes alignment independently of the coupling parameters, and alignment precedes aggregation under a stated condition on the tension-to-binding coupling (see Supplementary Information). Placing the observed onsets on the same stiffness axis, with the predicted sequence shaded (Fig. 5f), shows that the order in which $x_i$ and anisotropy rise matches the predicted PT1 to PT3 sequence. The comparison is ordinal: it tests the order of appearance, not a quantitative mapping between specific stiffness values and tensions, which we do not claim.

These measurements were made in senescent fibroblasts, whose metabolic and migratory activities are markedly reduced (14), yet the stepwise sequence is the same as in young primary cells and cell lines (2–5), which is consistent with a physical contribution to the reorganization. This preservation is suggestive rather than conclusive, however, since senescent cells retain active mechanotransduction, focal-adhesion signaling, and altered contractility; their reduced activity does not by itself isolate a purely physical mechanism. Even so, the persistence of the same sequence is consistent with the physically grounded account developed here and supports treating the hierarchy as a robust, stiffness-driven feature rather than one tied to a particular metabolic state.

## Discussion

Many studies have described how living cells reorganize their internal structures in response to mechanical cues by identifying the molecular pathways that mediate mechanotransduction (34–38). These studies have substantially advanced our understanding of how cells sense and respond to mechanical environments. However, how continuous mechanical cues bias the selection of distinct cytoskeletal architectures remains incompletely understood. To address this question, we developed a statistical-mechanical framework in which alternative cytoskeletal organizations are represented as ensembles of microscopic configurations, whose statistical preference is determined by the competition between configurational entropy and mechanically dependent interaction energies. Although statistical-mechanical approaches have previously been applied mainly to multicellular systems (39), here we extend this perspective to intracellular cytoskeletal organization within a single cell. This minimal framework identifies the physical principles sufficient to account for the experimentally observed sequence of cytoskeletal reorganization.

A key contribution of this study is that distinct cytoskeletal architectures can be understood as statistically selected states within a common statistical-mechanical framework. In this framework, configurational entropy and mechanically dependent interaction energies together determine how continuous mechanical inputs bias the selection of discrete cytoskeletal states. As substrate stiffness increases, changes in this balance progressively alter the statistical preference among alternative cytoskeletal architectures, giving rise to a sequence of threshold-dependent phase transitions that provides a physical basis for the experimentally observed progression from SF formation to directional alignment and lateral aggregation. Much like minimal models in statistical physics that extract essential behavior from complex systems, our approach captures the essential energetic relationships underlying these transitions without explicitly representing the full molecular complexity of the cytoskeleton. Importantly, expressing the three control parameters as increasing functions of a single cellular tension $F$, which rises monotonically with substrate stiffness, demonstrates that the observed sequence is not an artifact of parameter tuning. Instead, it emerges from a fixed hierarchy of thermodynamic thresholds ($F_1^{th} < F_2^{th} < F_3^{th}$), ensuring that fiber formation precedes directional alignment and lateral aggregation in a robust order. This predicted sequence is consistent with the experimental observations, in which actin filament formation, indicated by the initial increase in correlation length $\xi$, precedes directional alignment measured by anisotropy. Although bundle thickening was not quantified directly, the continued increase in correlation length on glass substrates is consistent with progressive bundle consolidation. Furthermore, the preservation of this sequence in senescent fibroblasts suggests that the proposed statistical-mechanical framework remains applicable despite substantial changes in cellular physiology.

The present framework also has implications beyond cytoskeletal organization itself. Failure to establish mechanically appropriate cytoskeletal architectures may alter downstream cellular responses to mechanical environments. Focal adhesions located at the termini of SFs convert mechanical tension into biochemical signals, and insufficient tension can lead to sustained activation of signaling pathways such as MAPK (40,41). Such mechanically dysregulated signaling has been associated with chronic inflammatory diseases, including atherosclerosis (42–44). While these molecular mechanisms have been investigated extensively (45,46), our framework provides a complementary systems-level perspective by suggesting that these signaling outcomes may arise from failures in the statistical selection of mechanically appropriate cytoskeletal architectures.

The sequence of structural transitions observed as cells spread on stiff substrates can be interpreted as a gradual process of internal mechanical maturation. This progression is driven by intracellular percolation (30), in which actin–myosin interactions increasingly assemble into SFs that support tension generation, leading to progressive intracellular stiffening. Within our framework, each statistically selected cytoskeletal state provides the mechanical basis for the emergence of the subsequent state, producing an ordered sequence of structural maturation. While cell cycle progression is canonically regulated by biochemical checkpoints such as cyclins and cyclin-dependent kinases, it is during the G1 phase, when cells attach and spread, that essential cytoskeletal reorganization occurs (3–5) to support entry into later phases (47). This stepwise maturation of intracellular structures, for which we have described the underlying physical conditions, may contribute to a series of mechanical checkpoints that align cytoskeletal organization with the timing required for proper cell cycle progression.

While the notion of mechanical checkpoints has appeared in previous studies, typically referring to threshold-dependent behaviors in cell polarization (48), focal adhesion dynamics (49,50), or mechanosensitive differentiation (51–53), these descriptions have remained largely phenomenological. Our framework provides a complementary interpretation by suggesting that such checkpoint-like behaviors can emerge naturally from the statistical selection of cytoskeletal architectures governed by configurational entropy and mechanically dependent interaction energies. In this view, mechanical checkpoints need not be regarded as independent regulatory events but may instead reflect physical constraints that govern intracellular architecture selection.

Given that SFs comprise more than one hundred associated proteins as revealed in our recent work (54), their full molecular diversity lies far beyond the scope of any minimal model. In addition, because cytoskeletal architecture varies across different regions of a cell, individual parts of the cytoplasm may exist at different stages of mechanical maturation during spreading. From this perspective, each statistically selected state in our framework should be interpreted as the dominant structural behavior within a local sub-system rather than the entire cytoplasm. The present model therefore focuses on the essential physical principles governing cytoskeletal organization while remaining readily extensible through modifications of the statistical weights or constraints that determine which filament configurations contribute most strongly to the ensemble. For example, regulation by actin-destabilizing molecules such as cofilin (27,28) can be incorporated without introducing explicit dynamical descriptions by biasing the statistical weighting toward shorter filament states. Supplementary Information presents one such example, demonstrating how cofilin-mediated actin destabilization can be incorporated within the same framework. These examples illustrate how additional molecular regulation can be naturally integrated into the statistical-mechanical framework through changes in the underlying free-energy landscape.

In recent years, the concept of liquid–liquid phase separation has provided a valuable framework for understanding the dynamic organization of membrane-less compartments in cells (55). More broadly, statistical mechanics has proven highly effective in explaining how complex biological organization can emerge from physically interpretable interactions. Our study shares a similar spirit but applies this perspective to mechanically regulated cytoskeletal organization rather than biomolecular condensation. By incorporating the catch-bond behavior of actin–myosin interactions, an essential feature of cellular mechanobiology (56), into a statistical-mechanical framework, we demonstrate how distinct cytoskeletal architectures can emerge through a hierarchy of threshold-dependent phase transitions.

While living cells are fundamentally nonequilibrium systems, structural changes such as cytoskeletal organization often occur under conditions where equilibrium-based approaches remain highly effective. This perspective is consistent with successful equilibrium-based analyses in biological physics, including protein folding (57), focal adhesion mechanosensitivity (58,59), and lipid and biomolecular phase separation (60,61), where equilibrium thermodynamics and statistical mechanics have provided valuable insights despite the intrinsically open nature of these systems. Our results suggest that a substantial component of stiffness-dependent cytoskeletal organization can be understood within an equilibrium statistical-mechanical framework, thereby providing a reference against which genuinely nonequilibrium contributions, such as temporally varying mechanical inputs or spatially heterogeneous energy landscapes, can be identified in future studies.

Our statistical-mechanical analysis establishes a framework for understanding how continuous mechanical cues drive the statistical selection of distinct cytoskeletal architectures. While our current formulation focuses on the actin cytoskeleton within a simplified two-dimensional lattice, it can be readily extended to include additional cellular components, higher-dimensional lattices, or curved SF arrangements, allowing more realistic representations of cellular architecture. Although these extensions will inevitably increase the complexity of the energetic and entropic descriptions, they retain the core statistical-mechanical framework. In particular, SFs of varying lengths or geometries can be accommodated by treating the present model as a local sub-system within a larger cellular domain. Although such detailed modeling is beyond the scope of the present study, the framework introduced here provides a foundation for future statistical-mechanical studies of cytoskeletal organization and may contribute to a broader physical understanding of intracellular architecture selection across diverse biological contexts.

## Data, Materials, and Software Availability

All study data are included in the article and/or Supplementary Information.

## Acknowledgments

We thank Intan Chairun Nisa for her contribution to the experiments. Y.U. is supported by the Japan Society for the Promotion of Science (JSPS). This study was supported in part by JSPS KAKENHI grants (23H04928 and 24KJ1649) and JST ACT-X (JPMJAX25L2).

## Author contributions

Y.U. and S.D. designed research; Y.U. performed research and analyzed data; and Y.U. and S.D. wrote the paper.

## Competing interests

The authors declare no competing interests.

## Figures

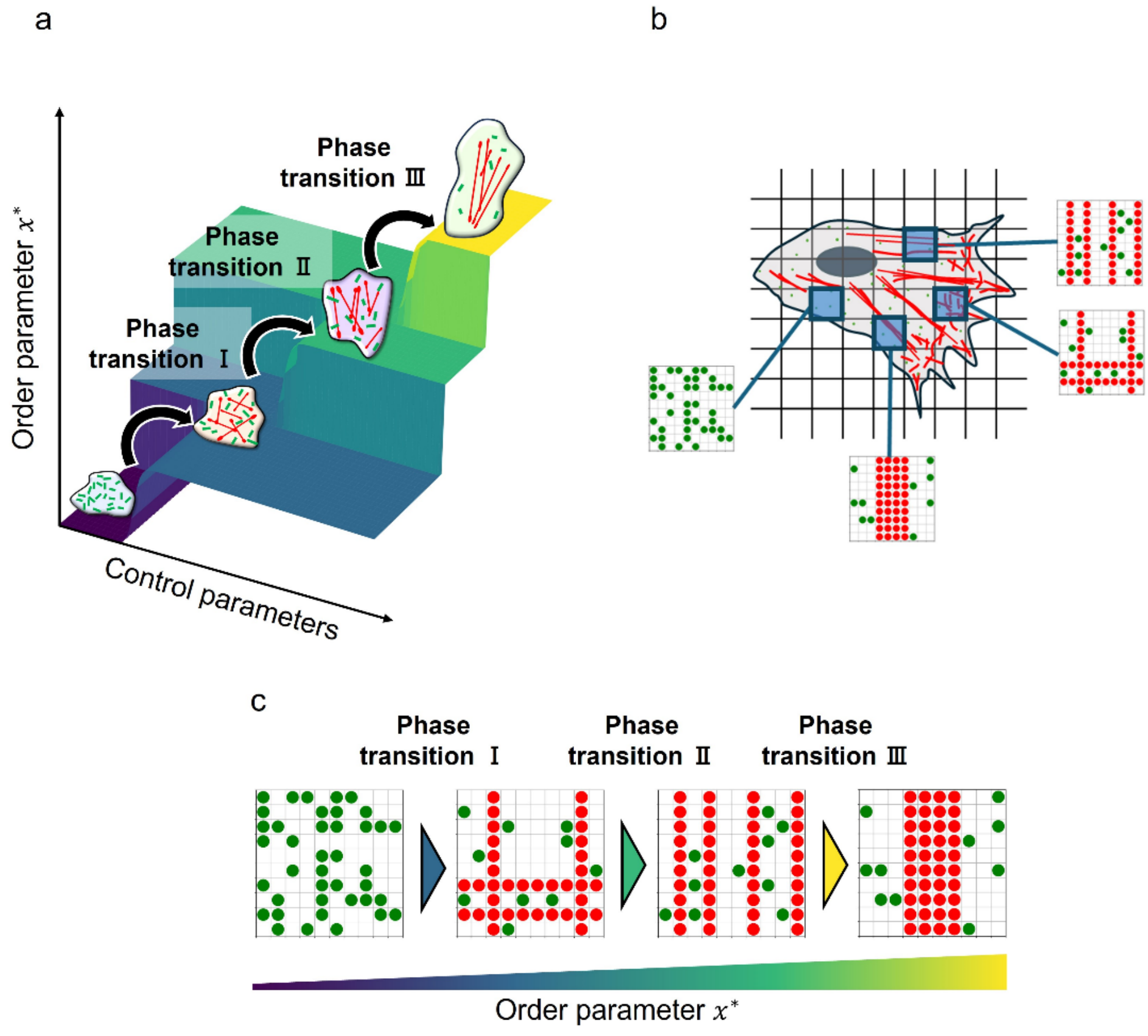

**Figure 1 Overview of stepwise cytoskeletal reorganization.**

(a) Intracellular actin structures undergo sequential phase transitions in response to control parameters such as substrate stiffness. (b) Coarse-graining of the cell into sub-systems: different cytoplasmic regions may occupy distinct structural states at the same time, and each region is treated as a single isolated lattice of side length a (right). The coexistence of different states in different regions is consistent with each region undergoing its own local transition. (c) Lattice-based cell models to analyze each transition: initially, randomly distributed actin particles (green) assemble into long filaments (red) (I), then align along a common direction (II), and eventually undergo interfiber coupling (III) with each transition characterized by a distinct order parameter. The order of each transition is determined in the following sections rather than assumed here.

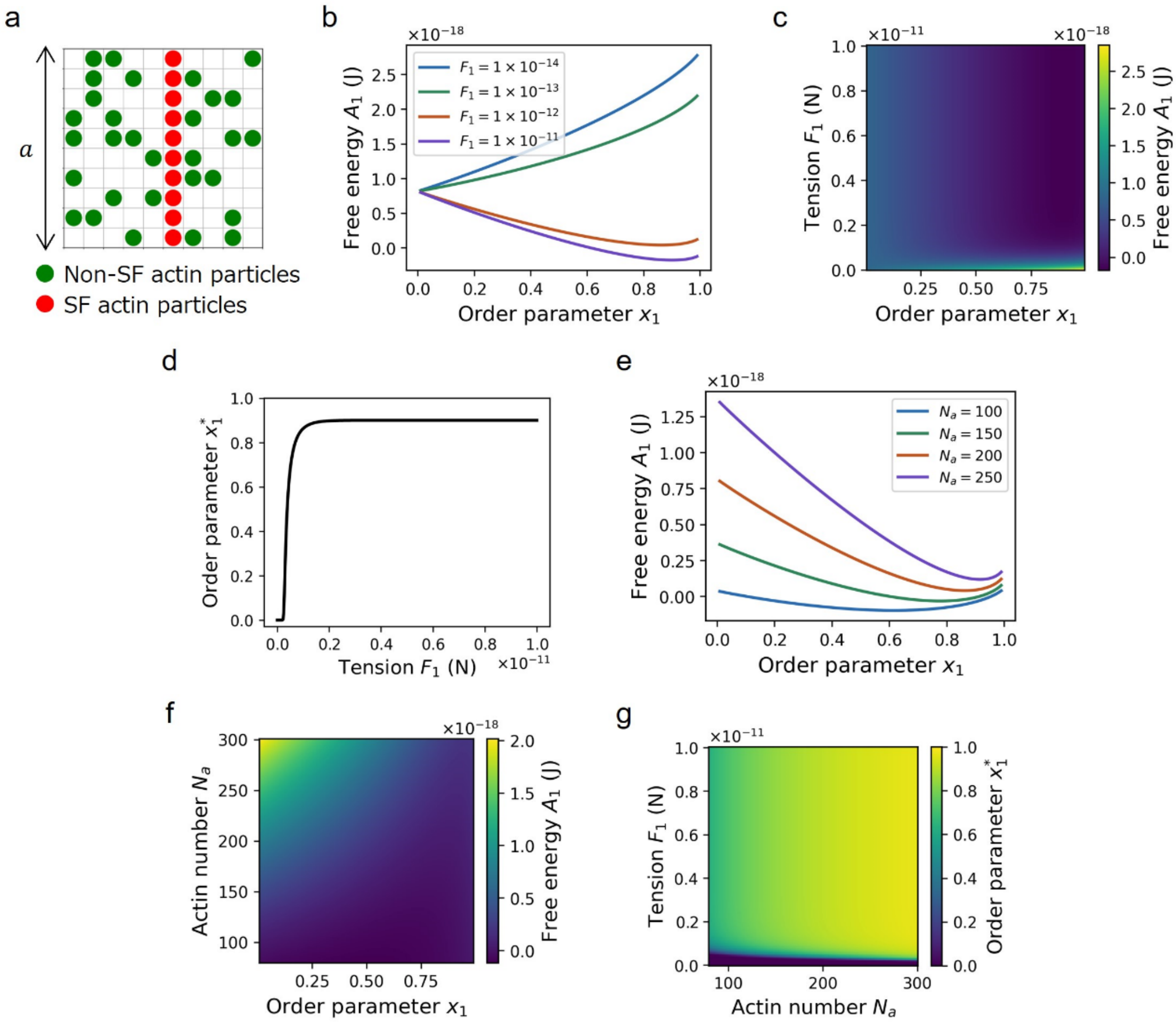

**Figure 2 Analysis of Phase transition I (PT1).**

(a) Schematic of a cell lattice model. Actin particles are classified as non-SF (green) or SF (red). (b) Free energy as a function of the order parameter and sustained tension. (c) Map of free energy as a function of the order parameter and sustained tension. (d) Order parameter that minimizes free energy as a function of sustained tension. (e) Free energy as a function of the order parameter and actin concentration. (f) Map of free energy as a function of the order parameter and actin concentration. (g) Critical order parameter as a function of the effective actin concentration and sustained tension.

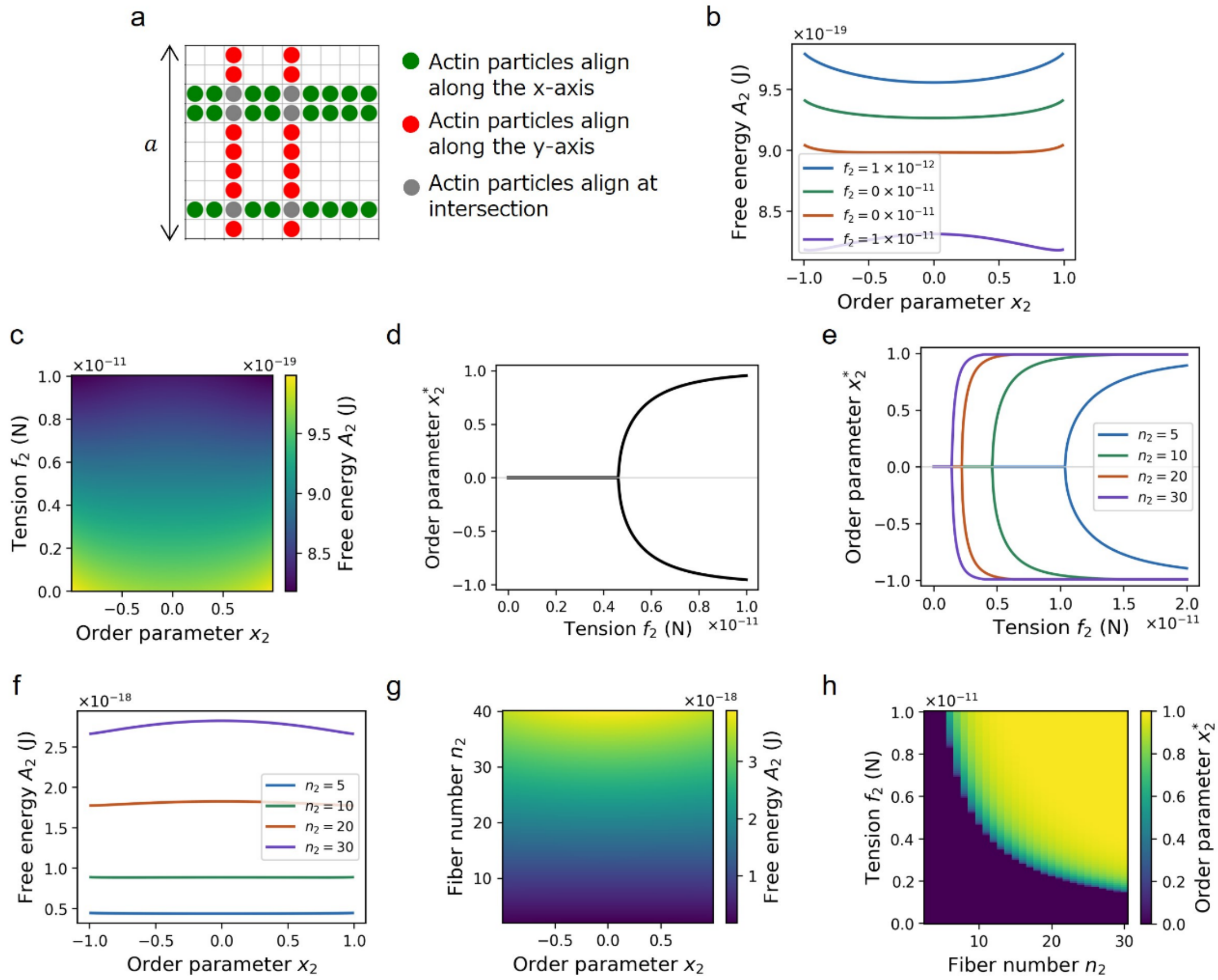

**Figure 3 Analysis of Phase transition II (PT2).**

(a) Schematic of a cell lattice model for PT2. Actin particles align along the $x$-axis (green) or $y$-axis (red) to form SFs, where tension generation is abolished at the intersection (gray). (b) Free energy as a function of the order parameter and sustained tension. (c) Map of free energy as a function of the order parameter and sustained tension. (d) Order parameter corresponding to the minimum free energy as a function of the sustained tension. (e) Order parameter corresponding to the minimum free energy as a function of the total number of actin particles and sustained tension. (f) Free energy as a function of the order parameter and the number of SFs. (g) Map of free energy as a function of the order parameter and the number of SFs. (h) Order parameter corresponding to the minimum free energy as a function of the number of SFs and sustained tension.

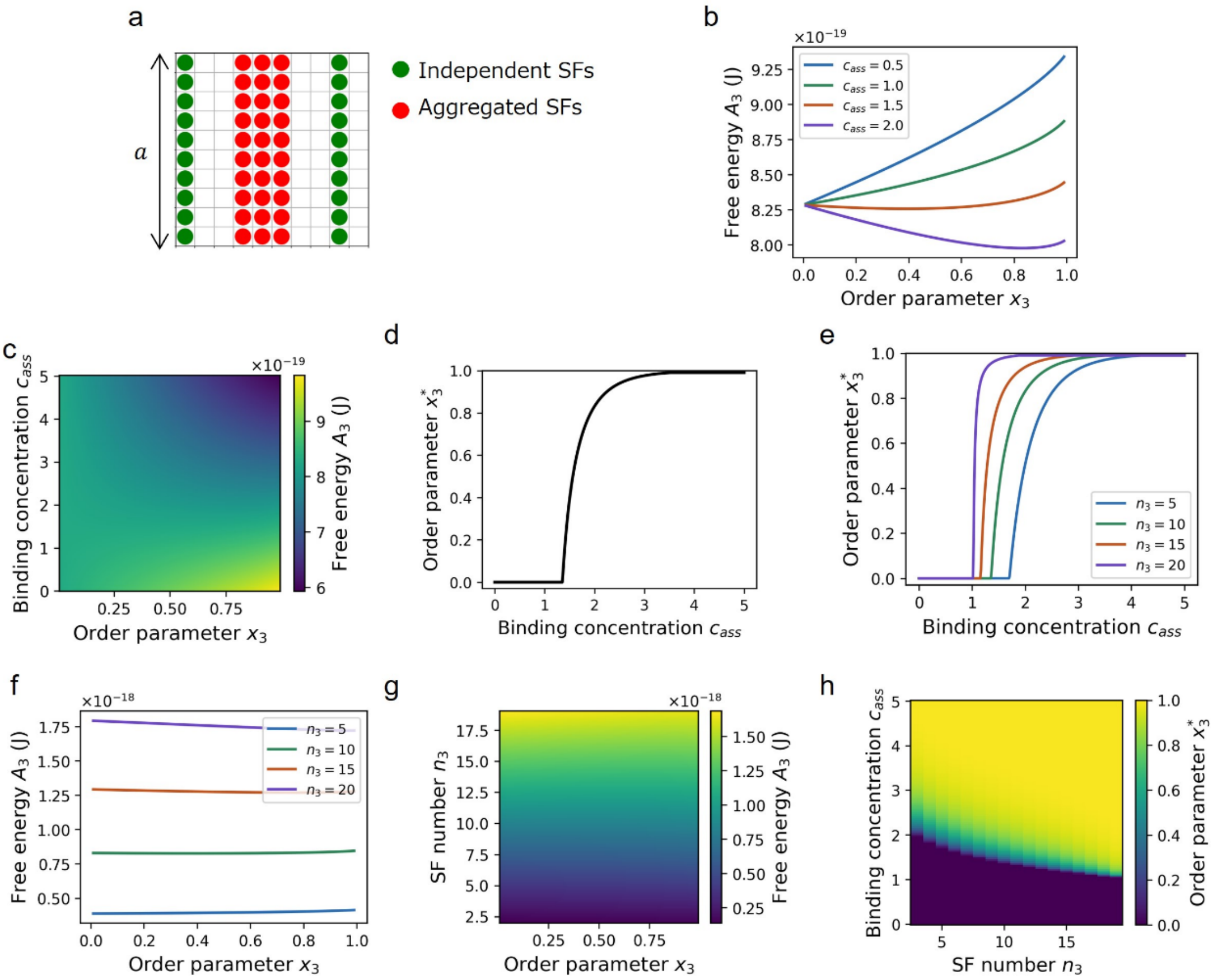

**Figure 4 Analysis of Phase transition III (PT3).**

(a) Schematic of a cell lattice model for PT3. Directionally aligned SFs are either distributed independently (green) or organized into aggregated structures (red). (b) Free energy as a function of the order parameter and effective binding protein concentration. (c) Map of free energy as a function of the order parameter and binding protein concentration. (d) Order parameter corresponding to the minimum free energy as a function of binding protein concentration. (e) Order parameter corresponding to the minimum free energy as a function of the total number of SFs and binding protein concentration. (f) Free energy as a function of the order parameter and the number of SFs. (g) Map of free energy as a function of the order parameter and the number of SFs. (h) Order parameter corresponding to the minimum free energy as a function of the number of SFs and binding protein concentration.

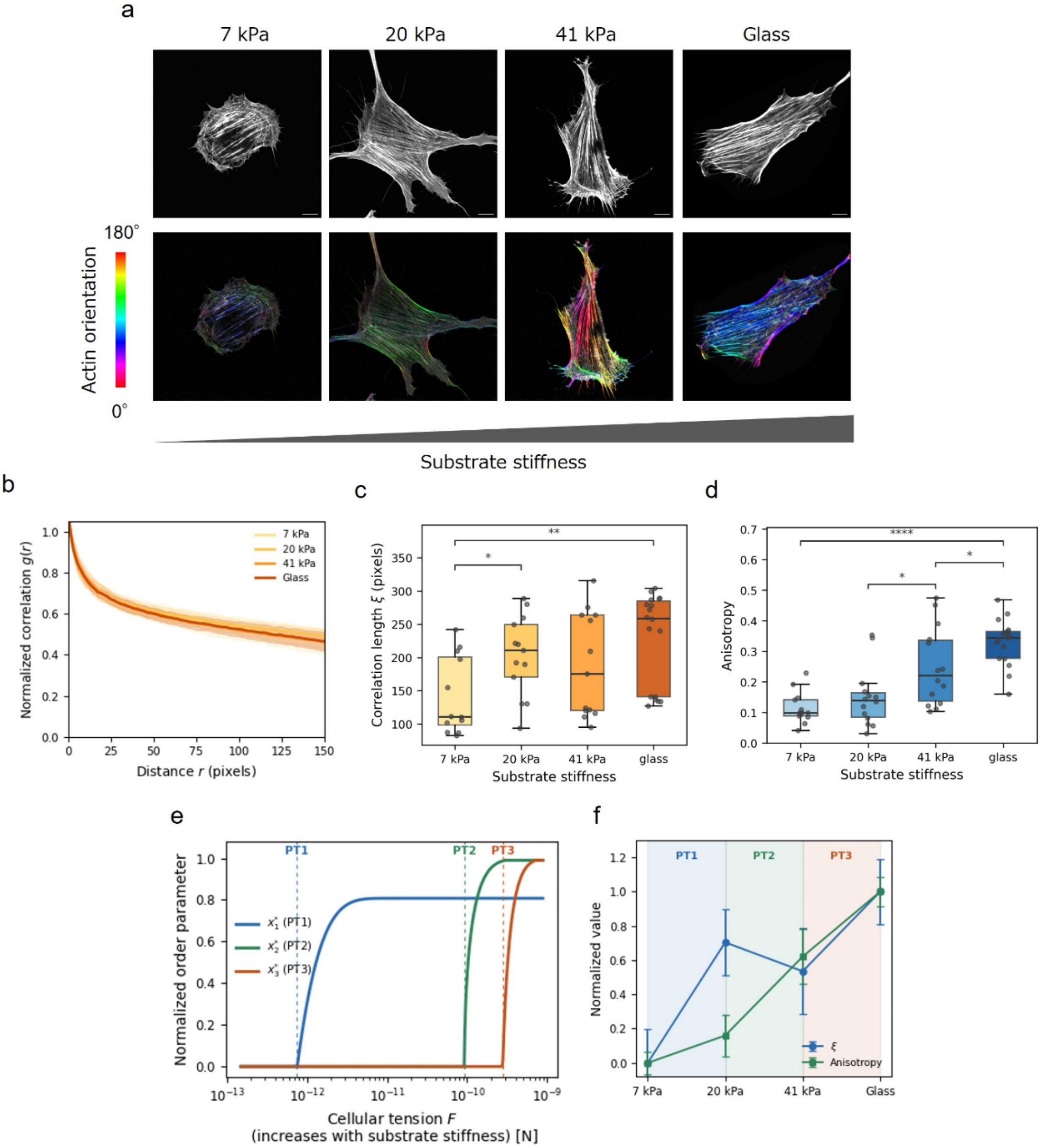

**Figure 5 Stiffness-dependent actin reorganization and comparison with the model.** (a) Representative F-actin images of senescent HFF-1 cells cultured on substrates of increasing stiffness (7, 20, and 41 kPa) and on glass (top), with corresponding local orientation maps highlighting filament orientation (bottom). (b) Normalized spatial autocorrelation functions of F-actin intensity (mean ± SD, $n$ = 10–15 cells per condition). (c) Correlation length $\xi$ extracted from exponential fits of the autocorrelation decay. (d) Anisotropy values quantifying directional alignment of actin structures. Asterisks denote statistical significance (*$p$ < 0.05, **$p$ < 0.01, ***$p$ < 0.001, and ****$p$ < 0.0001). (e) Model

order parameters $x_1^*$, $x_2^*$, $x_3^*$ on a single cellular-tension axis that increases with stiffness, showing the fixed order PT1, PT2, and PT3; dashed lines mark the thresholds $F_1^{th} < F_2^{th} < F_3^{th}$. (f) $x_i$ and anisotropy normalized to their ranges on the stiffness axis, with the predicted PT1, PT2, PT3 regions shaded; $x_i$ rises first (7 to 20 kPa) and anisotropy rises later (20 to 41 kPa), so the observed sequence is consistent with the model. Comparisons in (e, f) are ordinal and do not assert a quantitative stiffness-to-tension calibration.

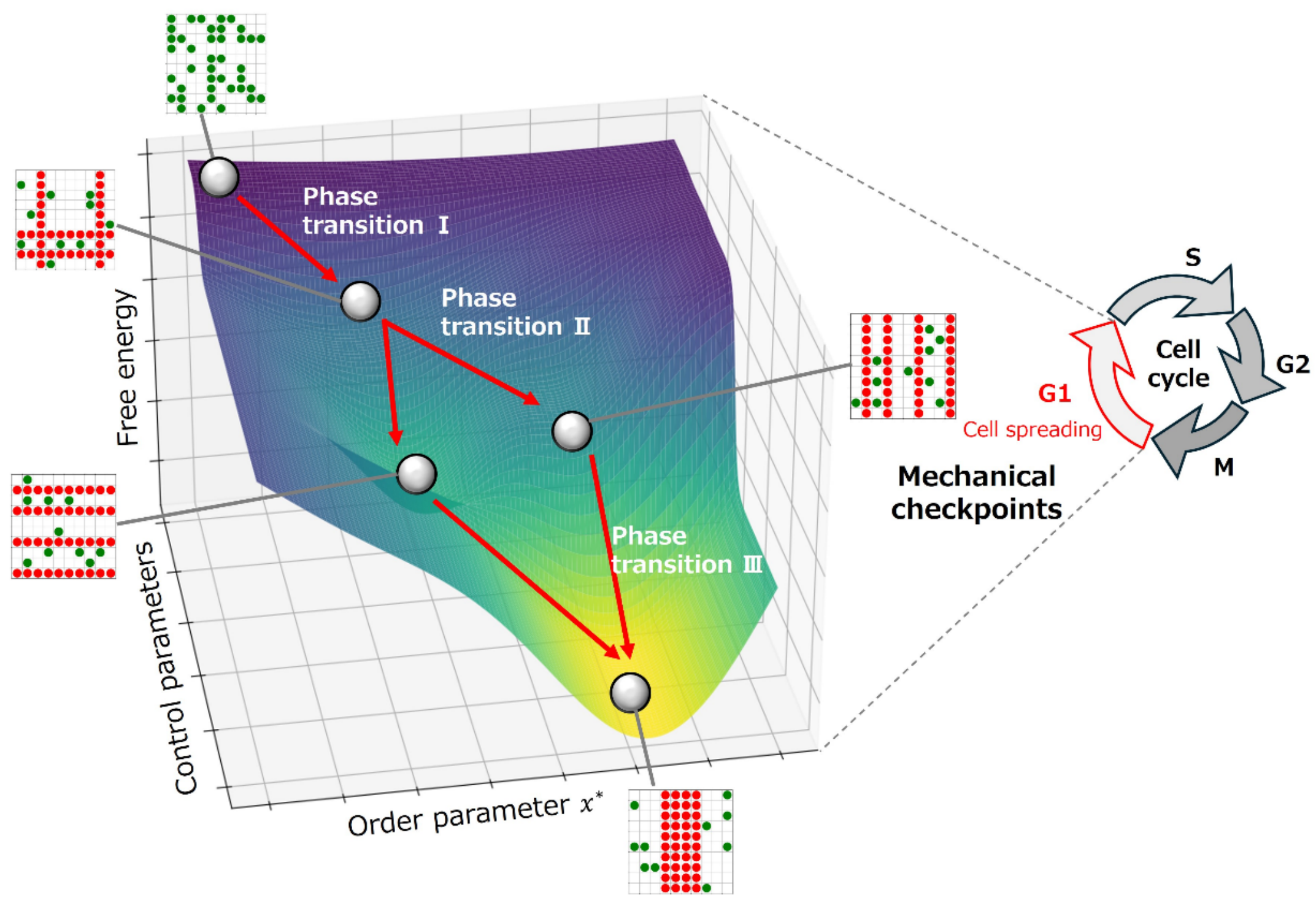

**Figure 6 Hierarchical free-energy landscape of mechanical cue-dependent intracellular transitions.**

A schematic free-energy surface illustrating three discrete cytoskeletal transitions (PT I–III) driven by increasing mechanical inputs such as substrate stiffness and the extent of cell spreading. Representative configurations highlight the structural regimes associated with each transition. The right panel presents a conceptual link between this hierarchy and mechanical checkpoint-like behaviors during G1-phase progression.

## Supporting Information

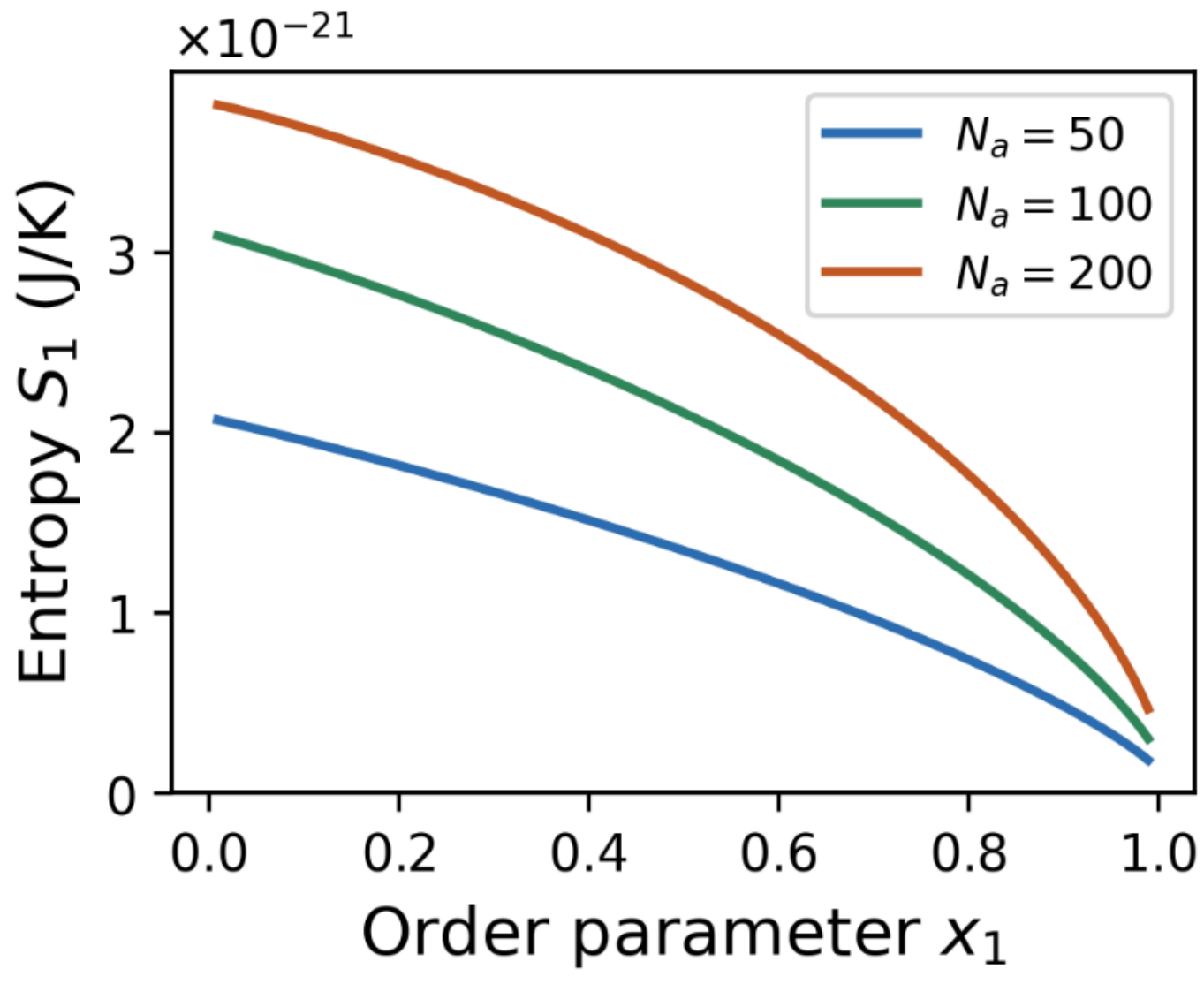

**Fig. S1 Entropy vs. order parameter for PT1.**

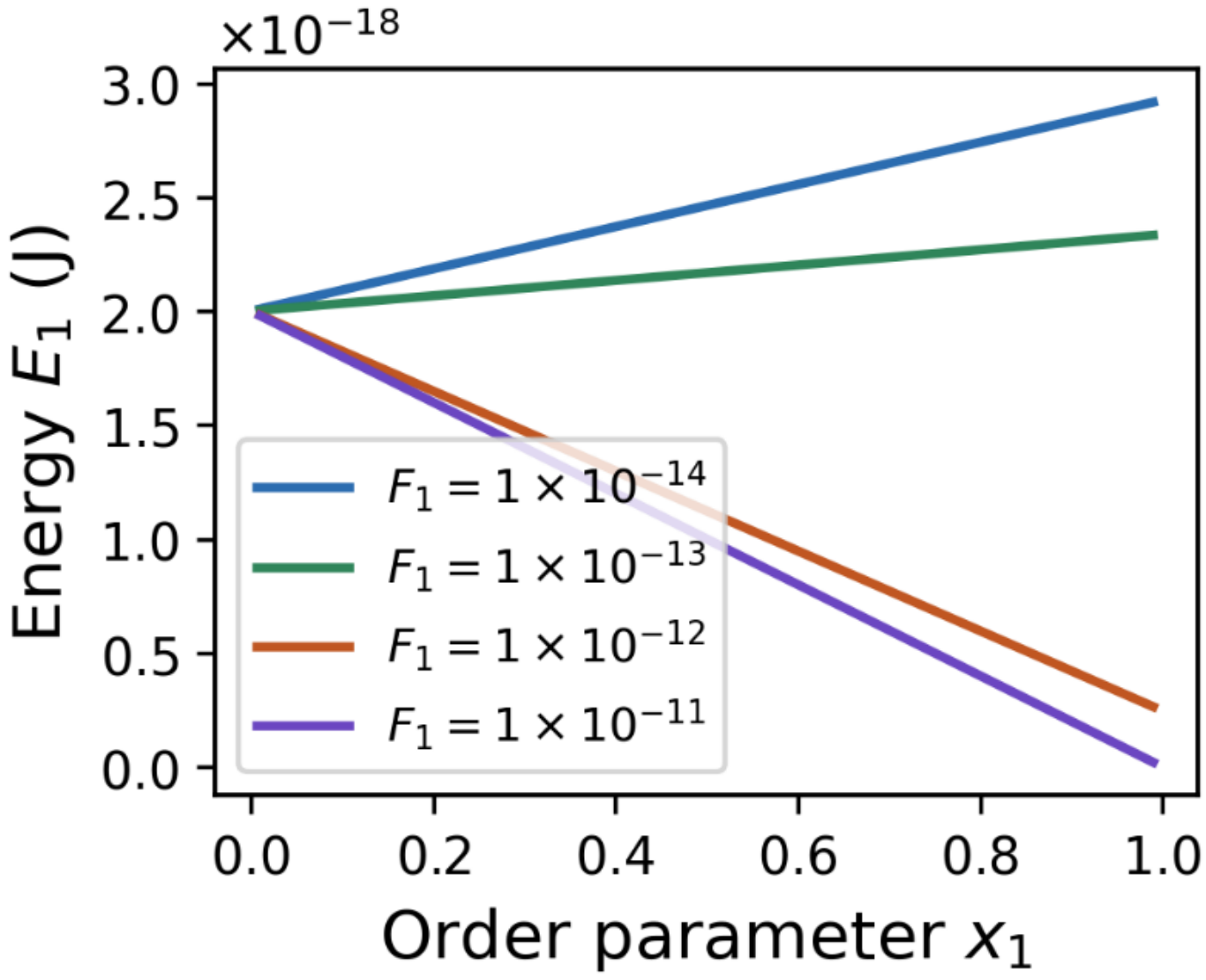

**Fig. S2 Energy vs. order parameter for PT1.**

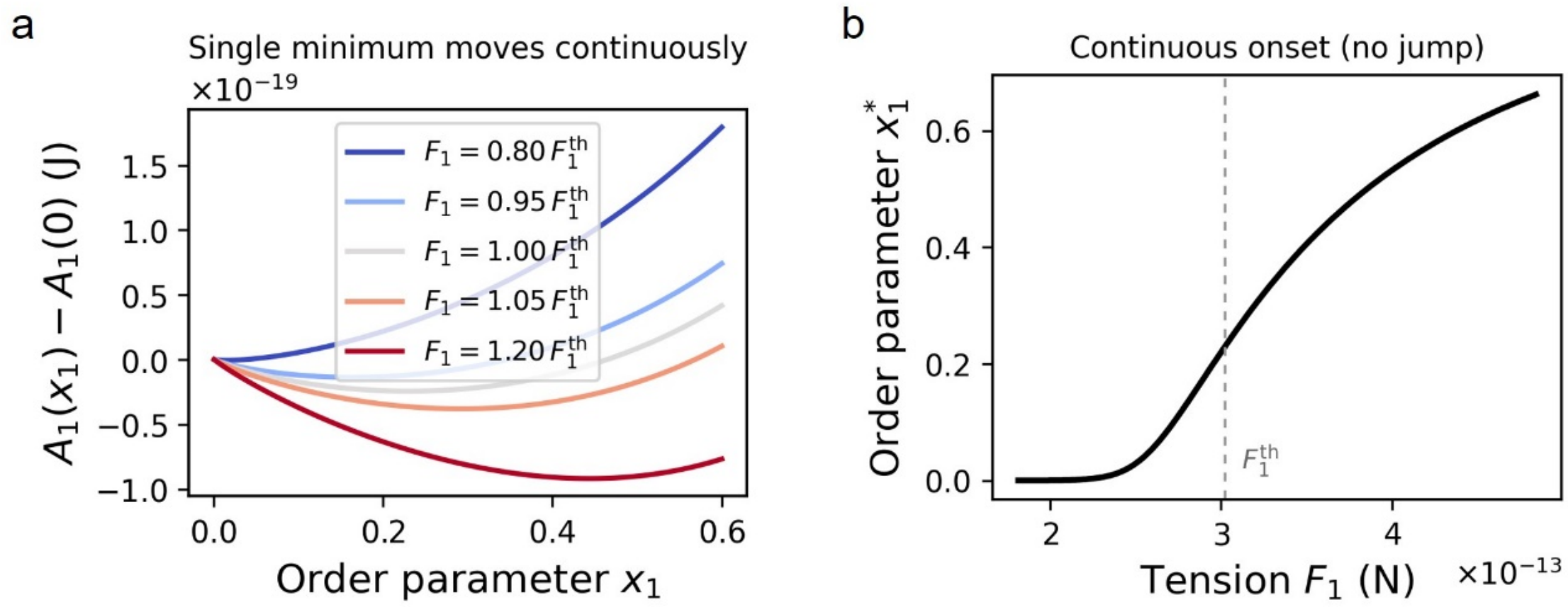

**Fig. S3 Characterization of the transition in PT1.**

(a) free-energy landscape $A_1(x_1)$ for tensions bracketing the threshold, showing a single minimum that moves continuously; (b) $x_1^*$ near the threshold, rising continuously from zero.

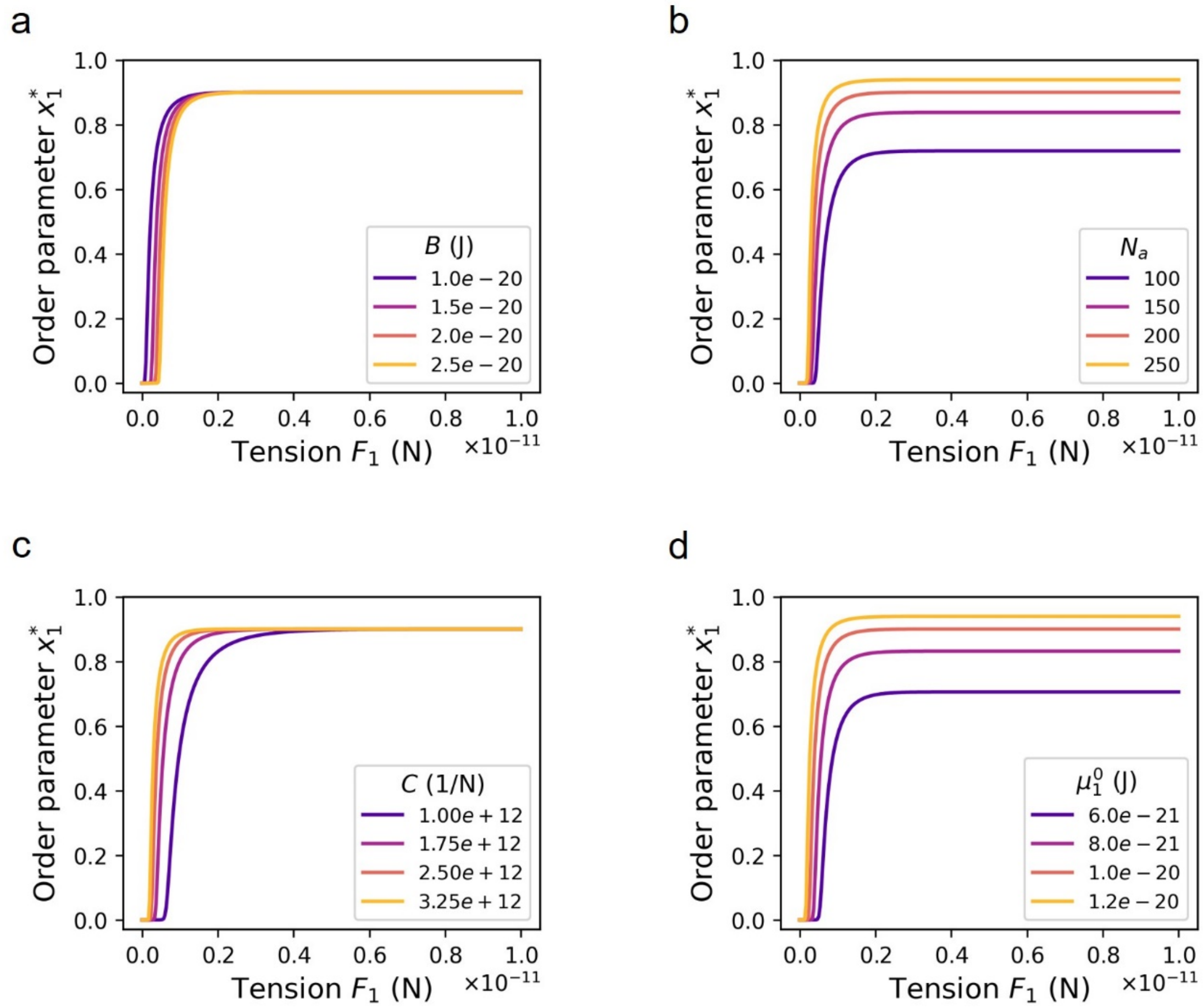

**Fig. S4 Parameter sensitivity for PT1.**

(a-d) $x_1^*$ as a function of tension while varying $B$ (a), $N_a$ (b), $C$ (c), and $\mu_1^0$ (d), showing that the transition persists and the threshold shifts smoothly.

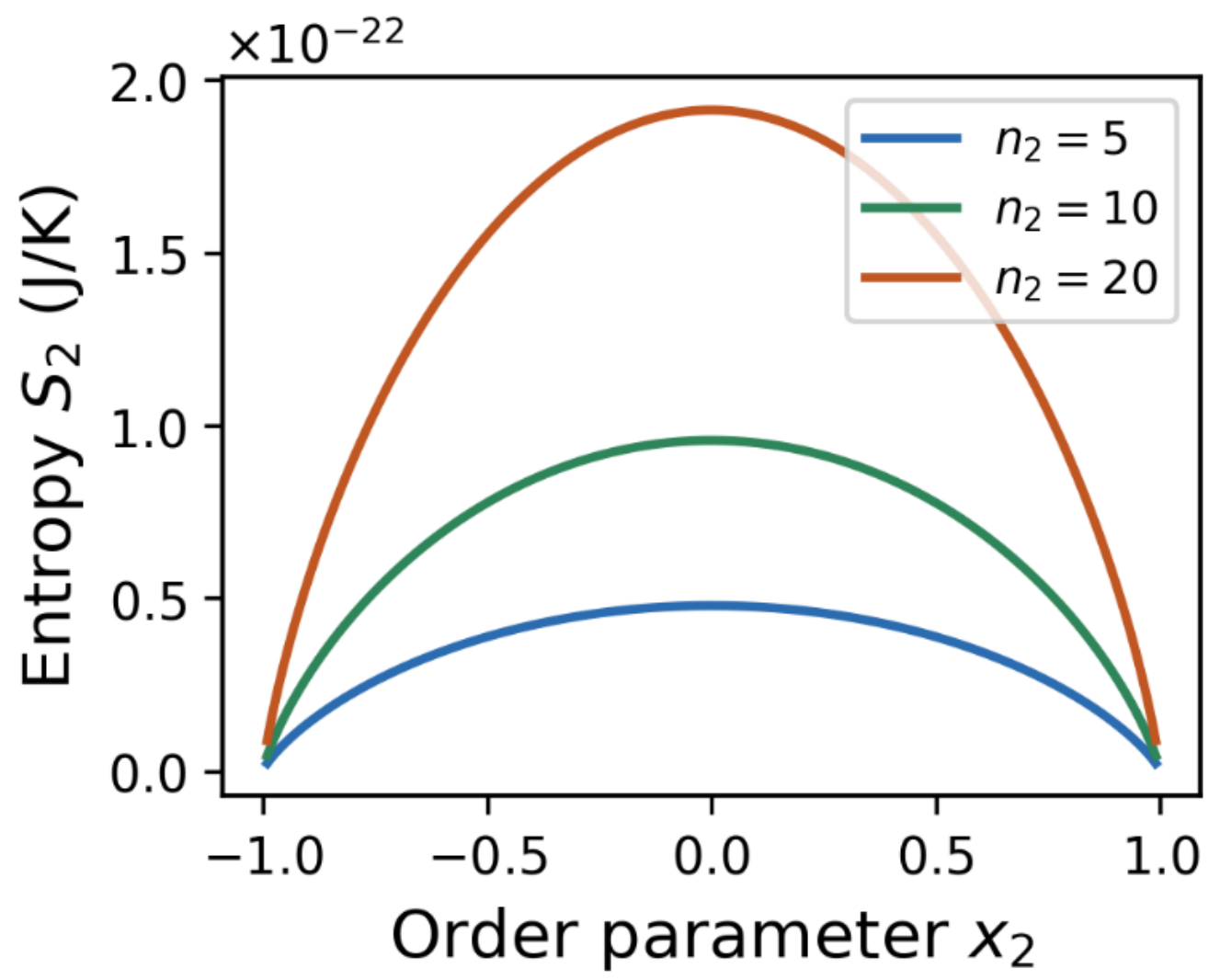

**Fig. S5 Entropy vs. order parameter for PT2.**

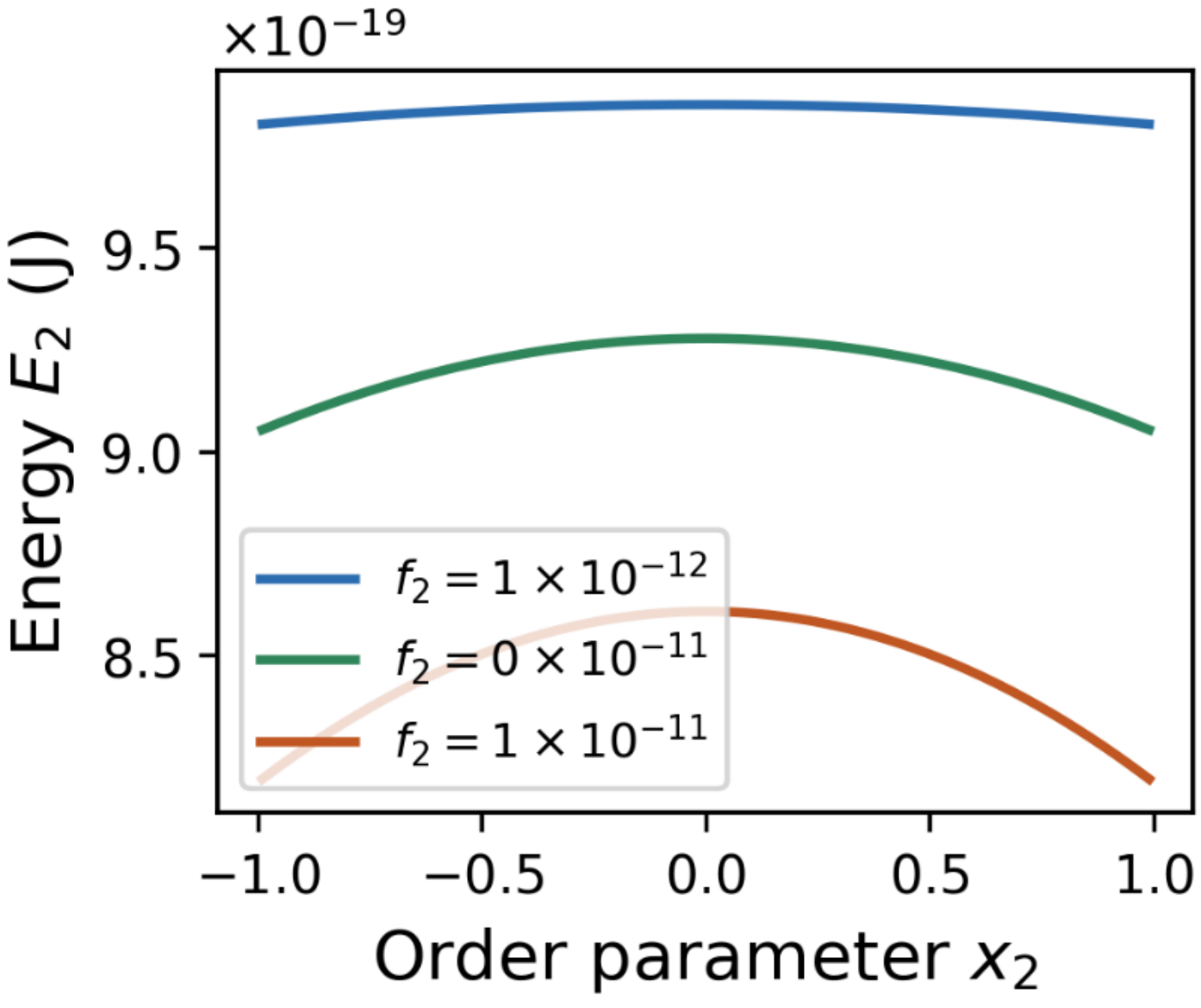

**Fig. S6 Energy vs. order parameter for PT2.**

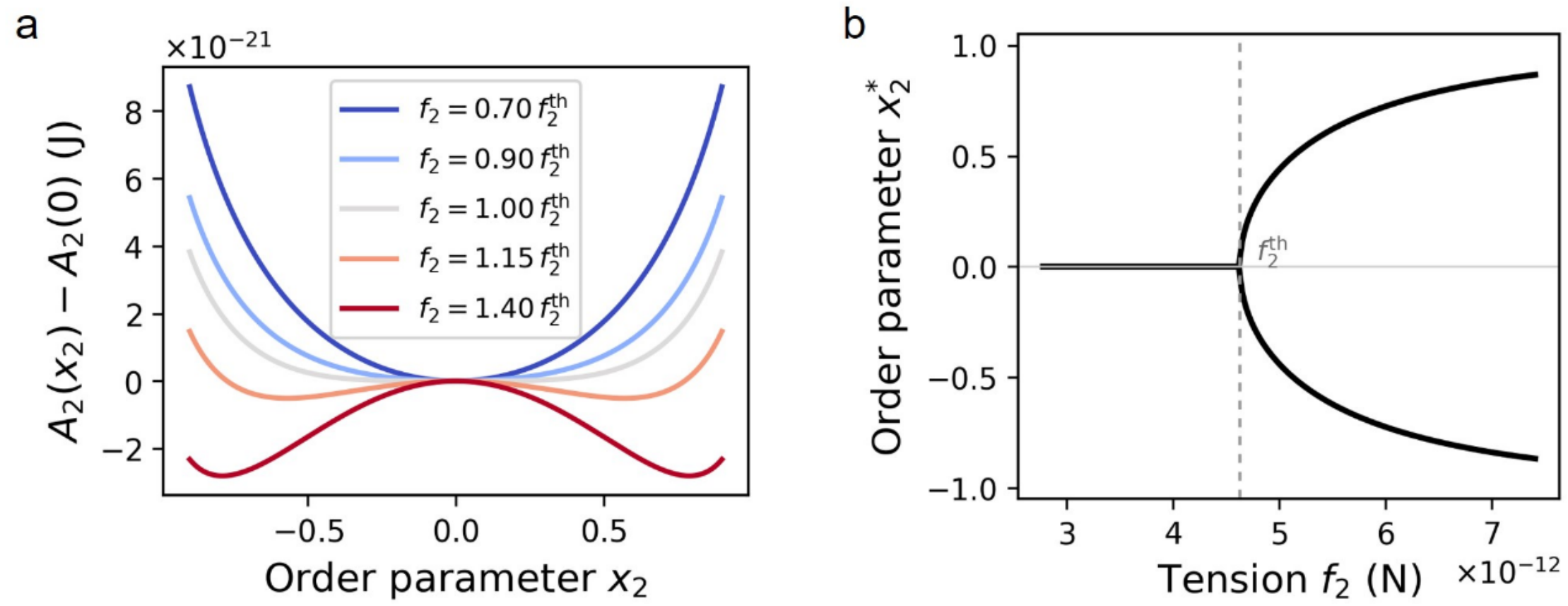

**Fig. S7 Characterization of the transition in PT2.**

(a) free-energy landscape $A_2(x_2)$ across the threshold, showing the single minimum at $x_2 = 0$ splitting into two symmetric minima. (b) The two branches of $x_2^*$ near the threshold, showing the pitchfork behavior.

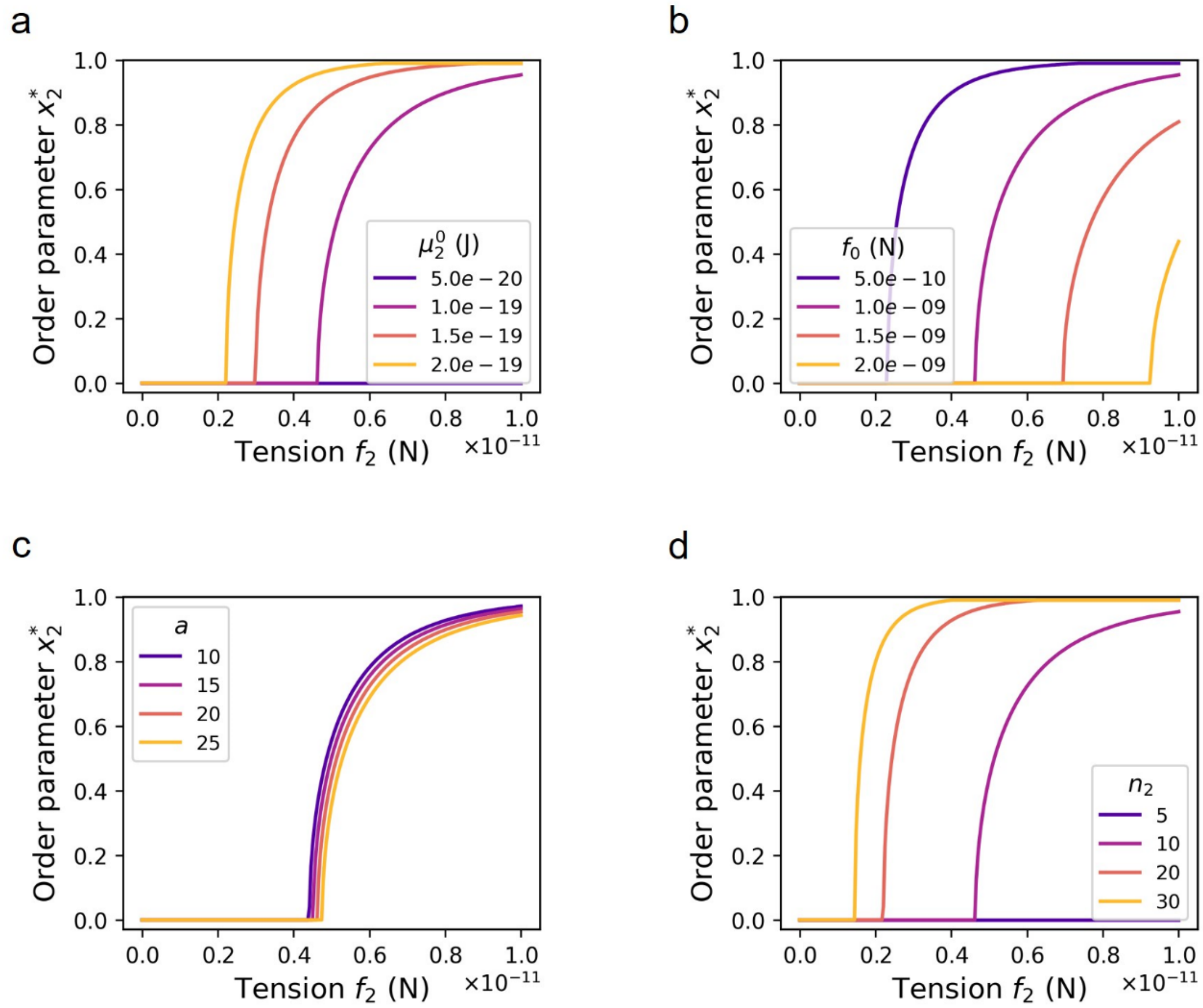

**Fig. S8 Parameter sensitivity for PT2.**

(a-d) $x_2^*$ as a function of tension while varying $\mu_2^0$ (a), $f_0$ (b), $a$ (c), and $n_2$ (d), showing that the transition persists and the threshold shifts smoothly.

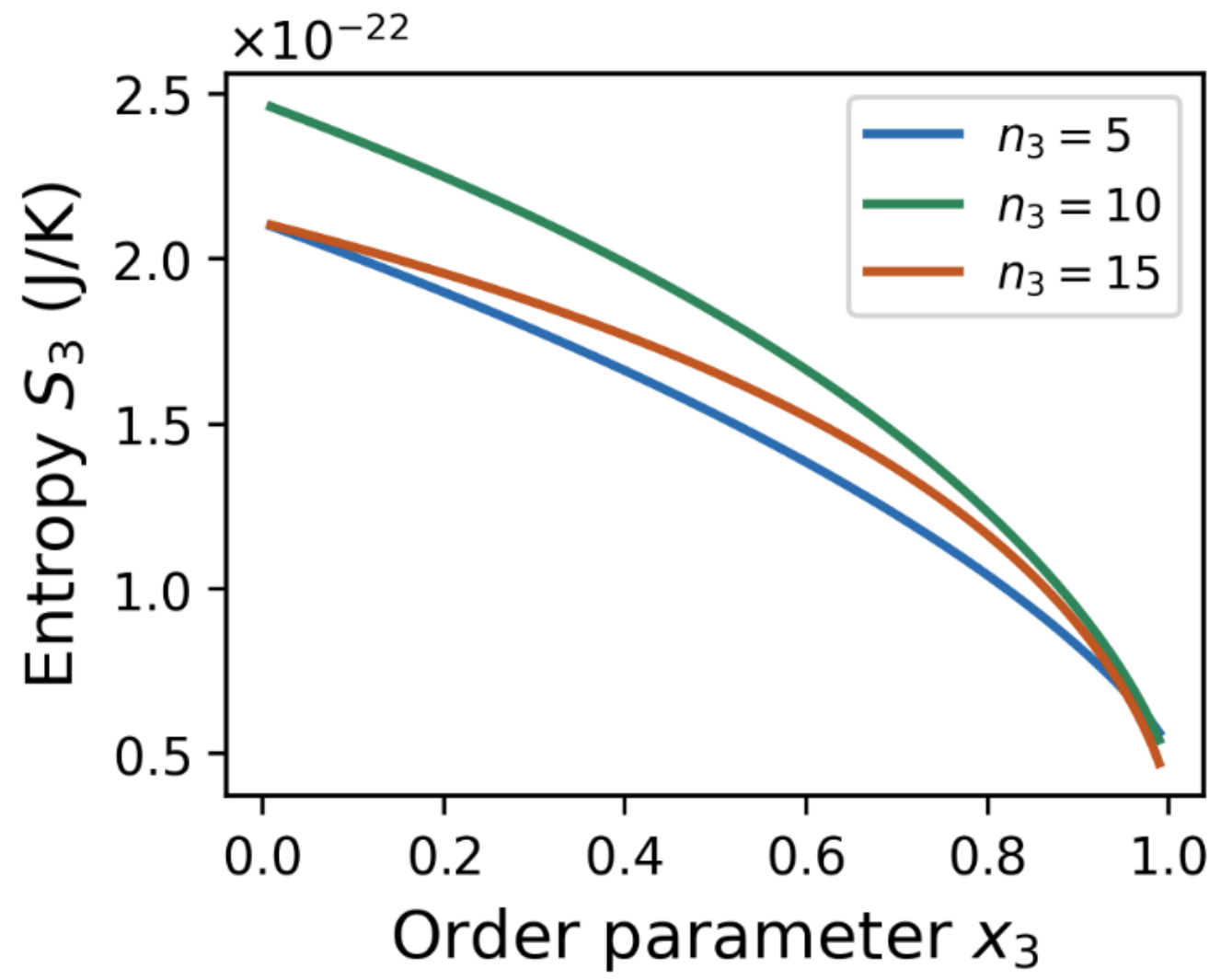

**Fig. S9 Entropy vs. order parameter for PT3.**

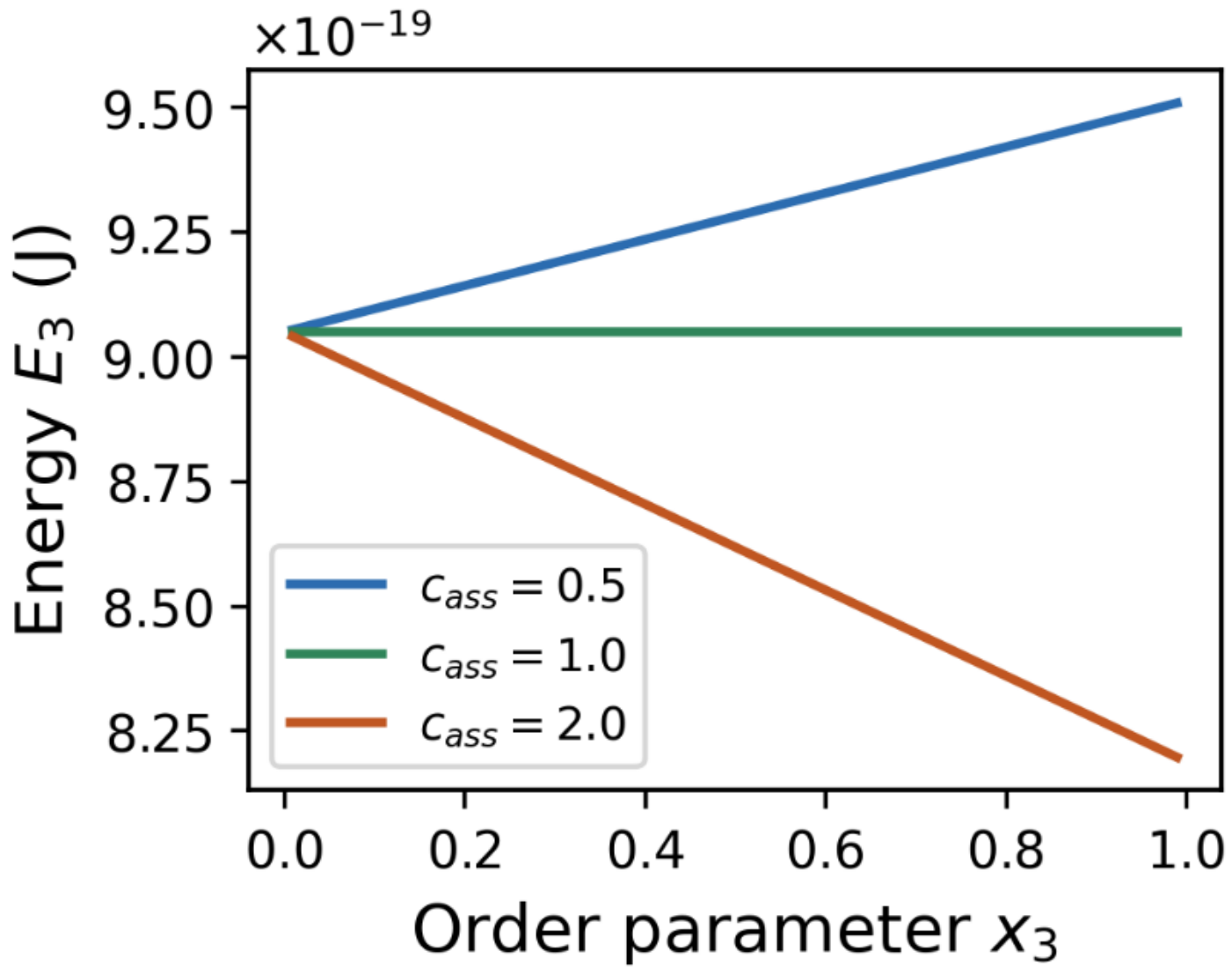

**Fig. S10 Energy vs. order parameter for PT3.**

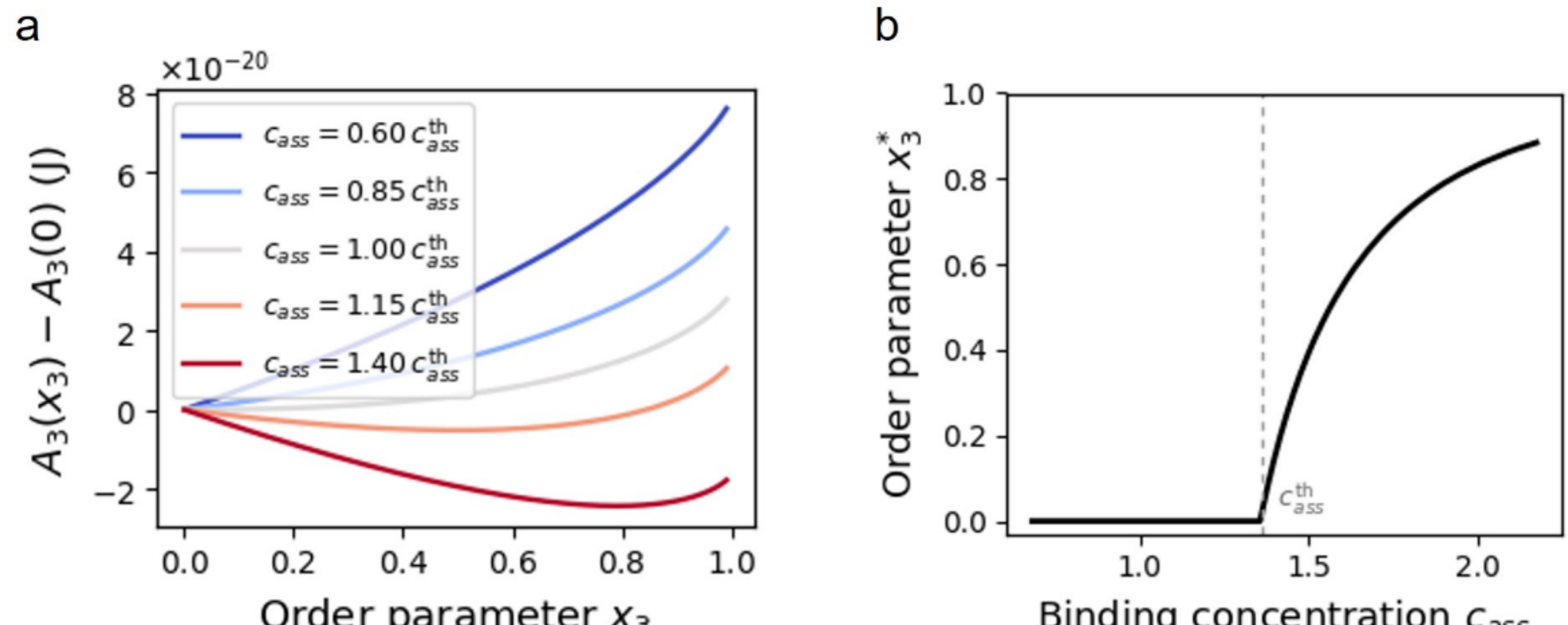

**Fig. S11 Characterization of the transition in PT3.**

(a) Free-energy landscape $A_3(x_3)$ for concentrations bracketing the threshold, showing a single minimum that moves continuously. (b) $x_3^*$ near the threshold, rising continuously from zero.

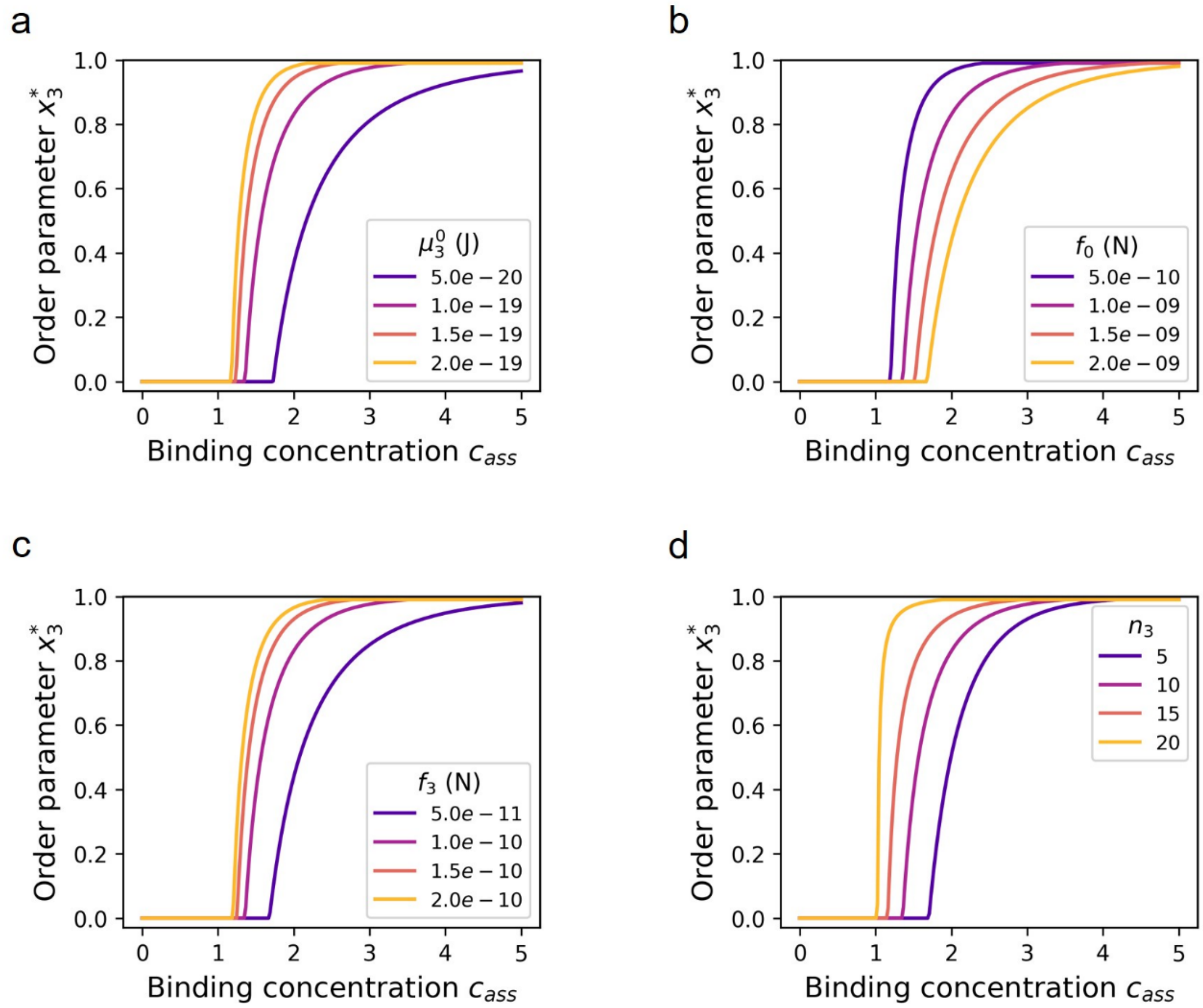

**Fig. S12 Parameter sensitivity for PT3.**

(a-d) $x_3^*$ as a function of binding protein concentration while varying $\mu_3^0$ (a), $f_0$ (b), $f_3$ (c), and $n_2$ (d).

## Order of the transitions

The order of each transition is determined from how the minimizing order parameter behaves as its control parameter crosses the threshold and from the shape of the free-energy landscape there. Across all three transitions, the onset is continuous, with no discontinuous jump and no hysteresis; PT2 additionally breaks a symmetry (pitchfork).

For PT1, expanding $A_1$ about $x_1 = 0$ using Eq. (1-10), we find that the onset of order occurs when $x_1 = 0$ ceases to be the minimum. As $A_1$ increases, $\mu_1^s = B\exp(-CF_1)$ decreases, so the energetic term $N_a(\mu_1^s - \mu_1^0)$ decreases while the entropic term favors small $x_1$. The minimizer leaves zero continuously (Eq. 1-11): just above threshold $x_1^*$ is small and positive and grows smoothly with $F_1$. Numerically, $A_1(x_1)$ keeps a single minimum that translates continuously from $x_1 = 0$ to positive $x_1$, with no second minimum and no hysteresis. PT1 is therefore a continuous (second-order-like) transition.

For PT2, because $A_2$ is symmetric, $x_2 = 0$ is always an extremum and the order is set by the curvature there. Below the threshold $f_2^*$ the curvature is positive and $x_2 = 0$ is a minimum; above it the curvature is negative and two symmetric minima appear at $x_2 = \pm x_2^*$. Near the threshold, $x_2^*$ grows continuously from zero as $\sqrt{f_2 - f_2^*}$, the standard signature of a continuous, second-order-like (pitchfork) transition. The two branches are physically equivalent and correspond to alignment along the two lattice axes.

For PT3, the minimizer $x_3^*$ (Eq. 3-9) equals zero below the threshold and rises continuously from zero above it; the single minimum of the free energy moves continuously from $x_3 = 0$ to positive $x_3$, with no second minimum and no hysteresis. PT3 is therefore a continuous transition.

# Table S1 Model parameters

| Symbol | Meaning | Value (default) | Units | Transition |
|---|---|---|---|---|
| $a$ | lattice side length | 20 | lattice units | all |
| $V = a^2$ | number of lattice sites (area) | 400 | sites | PT1 |
| $N_a$ | number of actin particles | 200 | particles | PT1 |
| $T$ | temperature | 309.5 | K | all |
| $k_b$ | Boltzmann constant | $1.38 \times 10^{-23}$ | J/K | all |
| $B$ | binding interaction energy at zero tension | $1.5 \times 10^{-20}$ | J | PT1 |
| $C$ | sensitivity of binding energy to tension | $2.5 \times 10^{12}$ | $N^{-1}$ | PT1 |
| $\mu_1^0$ | single-particle (intrinsic) energy | $1.0 \times 10^{-20}$ | J | PT1 |
| $F_1$ | sustained tension (PT1 control) | variable, ref. $1 \times 10^{-12}$ | N | PT1 |
| $x_1$ | order parameter (fraction of particles in SFs) | 0 to 1 | dimensionless | PT1 |
| $\mu_2^0$ | single-SF energy scale (PT2) | $1.0 \times 10^{-19}$ | J | PT2 |
| $F_2^0$ | tension normalizing factor (PT2) | $1.0 \times 10^{-9}$ | N | PT2 |
| $f_2$ | per-fiber tension (PT2 control) | variable | N | PT2 |
| $n_2$ | number of SFs (PT2) | 10 | fibers | PT2 |
| $x_2$ | order parameter (alignment) | -1 to 1 | dimensionless | PT2 |
| $\mu_3^0$ | single-SF energy scale (PT3) | $1.0 \times 10^{-19}$ | J | PT3 |
| $F_3^0$ | tension normalizing factor (PT3) | $1.0 \times 10^{-9}$ | N | PT3 |
| $f_3$ | per-fiber tension (PT3 control) | $1.0 \times 10^{-10}$ | N | PT3 |
| $c_{ass}$ | binding parameter of assembled state (PT3 control) | variable | dimensionless | PT3 |
| $n_3$ | number of SFs (PT3), constrained by $n_3 < a$ | 10 | fibers | PT3 |
| $x_3$ | order parameter (aggregation) | 0 to 1 | dimensionless | PT3 |

## Unified cellular tension and the ordering of the transitions

The three transitions were formulated with independent control parameters: the tension $F_1$ for PT1, the unit tension $f_2$ for PT2 (with per-fiber maximum tension $F_2 = af_2$), and the binding parameter $c_{ass}$ for PT3. In this formulation, the model does not fix the order in which the transitions occur. To obtain that order we use the shared physical driver invoked in each section, namely the cellular tension $F$, which increases monotonically with substrate stiffness and promotes actin–myosin stabilization through catch-bond behavior, and we express all three control parameters as increasing functions of $F$:

$$F_1 = F, \quad f_2 = \frac{F}{a}, \quad c_{ass} = \alpha F \qquad (S-1)$$

so that the effective tensions satisfy $F_1 = F_2 = F$, and $c_{ass}$ grows in proportion to tension with coupling constant $\alpha > 0$. The last relation follows the interpretation used in PT3, in which $c_{ass}$ reflects tension-enhanced binding-protein recruitment and focal-adhesion growth.

Each transition switches on when its control parameter crosses the threshold set by the corresponding free-energy balance. Converting these thresholds to the common tension axis gives the following. For PT1, the criterion (Eq. 1-12) with $\mu_1^s = B\exp(-CF_1)$ yields the threshold tension

$$F_{1,th} = -\frac{1}{C}\ln\frac{{\mu_1}^0 + k_b T\ln\frac{N_a}{V}}{B} \qquad (S-2)$$

Because $F_1$ and $F_2$ denote the same physical tension on this axis, PT1 precedes PT2 when $F_1^{th} < F_2^{th}$. For the model constants this holds by roughly two orders of magnitude, independently of $a$:

$$F_1^{th} \approx 7.6 \times 10^{-13}\, N < F_2^{th} = a f_2^{th} \approx 9.3 \times 10^{-11}\, \mathrm{N} \qquad (S-3)$$

where $f_2^{th}$ is the PT2 threshold. Formation therefore necessarily precedes alignment. For PT3, the threshold in $c_{ass}$ (from Eq. 3-10) maps to a threshold tension

$$F_3^{th} = \frac{c_{ass}^{th}}{\alpha} \qquad (S-4)$$

so alignment precedes aggregation, $F_2^{th} < F_3^{th}$, provided

$$\alpha < \frac{c_{ass}^{th}}{F_2^{th}} \approx 1.5 \times 10^{10} \qquad (S-5)$$

Under this single condition the thresholds satisfy $F_1^{th} < F_2^{th} < F_3^{th}$, so as the cellular tension $F$, and hence the substrate stiffness, increases, the transitions occur in the fixed order PT1, PT2, and PT3 (Fig. 5e). The first inequality is parameter-free; the second is an explicit condition on the coupling $\alpha$. The absolute position of PT3 depends on $\alpha$, but the ordering is preserved for any α below the bound. This mapping also yields a testable prediction: reducing the availability of cross-linking or adhesion-associated proteins would be expected to lower $\alpha$, raise $F_3^{th}$, and shift aggregation to higher stiffness, without affecting the formation and alignment thresholds.

## Extension of the PT1 model to actin filament destabilization by cofilin

Actin filament destabilization induced by actin-binding proteins such as cofilin can be incorporated into the PT1 framework as an effective increase in the energetic cost of the SF state. Here we present a minimal extension of the PT1 model to account for this effect.

Let the concentration of cofilin be denoted by [cof]. We extend Eq. (1-7) in the main text by modifying the effective interaction term as

$$\mu_{1s}^{s}(F_1) = B\exp(-CF_1) + \delta[\mathrm{cof}] \qquad (S-6)$$

where $\delta$ (> 0) represents the energetic contribution per unit cofilin concentration. The total energy is then given by

$$E_1 = \mu_1^0(N_a - L_s a) + \mu_{1s}^{s} L_s a, \qquad (S-7)$$

and the corresponding free energy, using Eq. (1-8) of the main text, becomes

$$A_{1S} = A_1 + \delta[\mathrm{cof}]L_s a. \qquad (\mathrm{S}-8)$$

To determine the phase transition threshold, we take the partial derivative of the free energy with respect to the order parameter, yielding

$$\frac{\partial A_{1S}}{\partial x_1} = \frac{\partial A_1}{\partial x_1} + \delta[\mathrm{cof}]N_a \qquad (\mathrm{S}-9)$$

Accordingly, the threshold condition for this extended model is

$$B\exp(-CF_1) + \delta[\mathrm{cof}] - \mu_1^0 < k_b T \ln\frac{N_a}{V}, \qquad (\mathrm{S}-10)$$

which plays a role analogous to that of Eq. (1–12) in the main text. Here, we assume that cofilin primarily affects the post-binding stability of actin filaments. Under this assumption, its effect is to reshape the energetic landscape by modifying the effective free energy of actin binding through changes in the local environment.

This extension of the model suggests experimentally testable predictions. For example, by inhibiting or overexpressing cofilin and comparing the resulting shifts in the phase-transition threshold relative to control conditions, one can probe not only the local chemical effects of cofilin on individual filaments but also its global physical influence on filament formation at the cellular scale, arising from competition between energetic destabilization and configurational entropy. Moreover, this framework is not limited to proteins with well-established roles in filament regulation. For actin-associated proteins with unclear functions, such as elongation factor eEF2, previously implicated in actin cross-linking (Liu et al., 2022), energetic contributions to filament organization can be inferred from shifts in the phase transition threshold. Thus, the model provides a statistical-mechanical framework in which mechanical and biochemical effects can be incorporated as changes in effective energetic parameters.

## Supplementary reference